\pdfoutput=1
\documentclass[aps,prl,twocolumn,longbibliography,superscriptaddress]{revtex4-2}
\usepackage{amsmath}

\usepackage{bm}
\usepackage{bbold}
\usepackage{amssymb}
\usepackage{graphicx}
\usepackage[ hidelinks = true, colorlinks = true, citecolor = blue, urlcolor = blue, linkcolor = blue]{hyperref}
\usepackage{verbatim}
\usepackage{color}
\usepackage[capitalise]{cleveref}
\usepackage[normalem]{ulem}
\usepackage{dsfont}
\usepackage{xcolor}
\usepackage{physics}
\usepackage{placeins}

\usepackage{svg}

\allowdisplaybreaks[3]

\usepackage{dsfont}

\newcommand{\atstrike}[1]{%
  \ifmmode
    \textcolor{red!80!black}{\cancel{#1}}%
  \else
    \textcolor{red!80!black}{\sout{#1}}%
  \fi
}

\newcommand{\ats}[1]{\textcolor{red!80!black}
{\ifmmode\text{\cancel{\ensuremath{#1}}}\else\sout{#1}\fi}} 

\usepackage{accents}
\newlength{\dhatheight}

\usepackage[normalem]{ulem} 

\begin{document}

\title{Adiabatic preparation of a fractional quantum Hall fluid by coherently pumping atoms from a Bose-Einstein condensate}

\author{Alberto Tabarelli de Fatis}
\affiliation{Pitaevskii BEC Center, INO-CNR and Dipartimento di Fisica, Universit\`{a} di Trento, Trento, Italy}

\author{Christof Weitenberg}
\affiliation{Department of Physics, TU Dortmund University, Dortmund, Germany}

\author{Alexander Schnell}
\affiliation{Technische Universität Berlin, Institut für Physik und Astronomie, Berlin, Germany}

\author{Andr\'e Eckardt}
\affiliation{Technische Universität Berlin, Institut für Physik und Astronomie, Berlin, Germany}

\author{Iacopo Carusotto}
\affiliation{Pitaevskii BEC Center, INO-CNR and Dipartimento di Fisica, Universit\`{a} di Trento, Trento, Italy}

\date{\today}

\begin{abstract}
We propose a protocol to adiabatically prepare a many-particle fractional quantum Hall fluid of bosonic ultracold atoms exploiting a time-dependent coherent coupling of a strongly interacting atomic state with a large dilute Bose-Einstein condensate. Starting from an empty cloud, atoms with well-defined angular momentum are coherently pumped into the fluid by Raman beams with a Laguerre-Gauss profile. 
Compared to number-conserving schemes which rely on finite-size-induced gaps, we identify an adiabatic path in the Fock space which avoids crossing fixed-particle-number phase transitions  and maintains a sizable and externally controlled adiabatic gap open at all times. The efficiency of our preparation protocol is numerically assessed for typical experimental parameters up to particle numbers that largely exceed the experimental state-of-the-art. The crucial advantage of including an anharmonic confinement is finally highlighted.

\end{abstract}

\maketitle

\paragraph{Introduction -}
Fractional quantum Hall (FQH) systems exhibit extremely rich and interesting properties arising from the interplay of strong interactions and the frustration induced by strong magnetic fields. They support topologically protected, quantized charge transport along their edges~\cite{PhysRevLett.49.405, Wen01101995}, and have been predicted to display (non)-abelian anyonic excitations with complex statistical properties~\cite{PhysRevLett.52.1583,PhysRevLett.53.722,MOORE1991362}, which are promising for applications in topological quantum computation and error correction~\cite{RevModPhys.80.1083}. While FQH physics was first observed in two-dimensional electron gases~\cite{PhysRevLett.45.494,PhysRevLett.48.1559}, realizing such states in a well-controlled and tunable platform is a long-standing goal. This would help better understand their strongly correlated nature and allow to fully unchain the potential of their quantum entanglement for different tasks in quantum science and technology.

In the last years, first experimental realizations of FQH fluids in both photonic~\cite{clark2020observation, doi:10.1126/science.ado3912} and atomic~\cite{gemelke2010rotatingfewbodyatomicsystems,leonard2023realization,PhysRevLett.133.253401, kwan2026pfaffianquantumhallstate} platforms have been reported, but these experiments remain so far limited to a maximum of three particles. In particular, the preparation of larger FQH states of ultracold atoms is hindered by the difficulty of directly cooling in the FQH regime. Typical workarounds involve the preparation of some other easy-to-cool state, followed by an adiabatic change of the system parameters in order to enter the FQH regime. This implies crossing a phase transition, which limits the achievable system sizes as the energy gap quickly decreases for growing particle numbers. Schemes that do not suffer from this difficulty have already been proposed for gases of photons~\cite{PhysRevLett.108.206809,clark2020observation,dutta2018coherent, umucalilar2017generation, kurilovich2022stabilizing}, and of Rydberg excitations~\cite{PRXQuantum.3.030302}.

\begin{figure}[h]
    \centering
    \includegraphics[width=\linewidth]{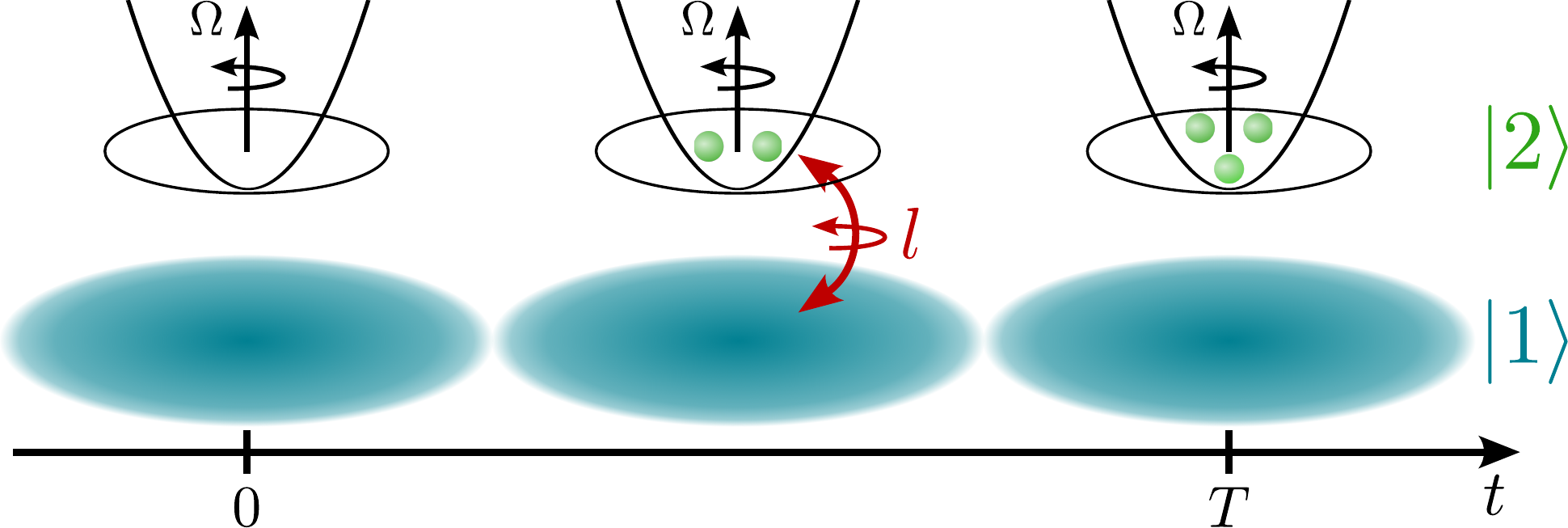}
    \caption{Proposed scheme for the generation of an FQH fluid of ultracold atoms. The FQH system under study is formed of atoms in a strongly interacting state $\ket{2}$ close to a Feshbach resonance (see Appendix), tightly trapped in a two-dimensional geometry; in the frame rotating at frequency $\Omega$, the atoms experience an effective synthetic magnetic field. This FQH system is coherently coupled via two Raman beams with a Laguerre-Gauss profile to a weakly interacting condensate $\ket{1}$ that acts as a particle reservoir, exchanging an angular momentum $l$ per particle. The FQH is initially empty, and is adiabatically pumped to the desired state in a time $T$.}
    \label{fig:scheme}
\end{figure}

\paragraph{Proposal -}
We propose a novel method for generating bosonic Laughlin states of atoms via an adiabatic protocol in a larger space where the number of particles is allowed to vary by an external coherent pump. Our scheme, sketched in \cref{fig:scheme}, is based on the coherent coupling of the system to a reservoir of Bose-Einstein condensed (BEC) atoms -- distinguished from the system by a different internal spin state -- via an angular-momentum-selective mechanism based on, e.g., a pair of Raman beams with a Laguerre-Gauss profile. 
By adiabatically varying the strength and detuning of the coherent coupling, the initial vacuum state can be converted into the target Laughlin state.

In our setup, the energy gap protecting the evolution is not fixed by the many-body physics of the fractional quantum Hall system, as in typical number-conserving schemes, but is externally controlled by the amplitude of the coherent coupling, which allows to keep a sizable gap throughout the whole adiabatic evolution. Laguerre-Gauss Raman coupling was also considered in Ref.~\cite{Zhang2016}, but coupling each angular momentum transfer to a spin flip fundamentally limits the target angular momentum to the size of the atomic spin manifold. Unlike~\cite{PRXQuantum.3.030302}, we assume a pump carrying angular momentum, which avoids the rotation of an external potential~\cite{roncaglia2011rotating,PhysRevA.70.053612,PhysRevA.103.063325} to insert angular momentum. In contrast to photonic setups~\cite{clark2020observation, PhysRevLett.108.206809}, losses do not dominate the evolution of the atomic system, allowing to follow the coherent dynamics for long times. Furthermore, our proposal of pumping particles from a BEC into the system has the advantage of not relying on the initial preparation of an exact atom number, which limited early experiments~\cite{gemelke2010rotatingfewbodyatomicsystems}. As compared to Ref.~\cite{PhysRevLett.113.155301,dutta2018coherent}, we propose to directly generate the $N$-particle Laughlin state in a single adiabatic ramp, avoiding the sequential accumulation of infidelity in multi-step protocols. 

The concept of the proposed scheme is illustrated in \cref{fig:scheme}. We consider atoms with two internal states $\ket{1},\ket{2}$, split by an energy $\hbar \omega_{1,2}$. These states define reservoir and system, respectively. The $\ket{1}$ state is weakly interacting and filled by a BEC with a macroscopic number of particles. The strongly interacting $\ket{2}$ state is trapped in a two-dimensional geometry and initially empty. Finally, an external coherent electromagnetic field coupling the two states is used to coherently transfer atoms from $\ket{1}$ to $\ket{2}$ and, in this way, generate a FQH fluid in the $\ket{2}$ state. This configuration can be realized with several atomic species, but for concreteness, we consider typical parameters for $^7$Li atoms (see End Matter).

\paragraph{Theoretical modeling -}
The dynamics of the atoms in state $\ket{2}$ can be described in a frame rotating at a constant angular frequency $\mathbf{\Omega} =-\Omega \mathbf{e_z}$, in which the Hamiltonian reads~\cite{cooper2008rapidly}:
\begin{equation}
    \begin{split}
     \hat{H}_{\ket{2}} = \int d^2 r & \hat{\Psi}^\dagger(\mathbf{r})\left[\frac{(\mathbf{\hat{p}}+\mathbf{A})^2}{2m}+V(r) + \hbar \omega_2\right]\hat{\Psi}(\mathbf{r})\\
    & +\frac{g}{2}\hat{\Psi}^\dagger(\mathbf{r}) \hat{\Psi}^\dagger(\mathbf{r}) \hat{\Psi}(\mathbf{r}) \hat{\Psi}(\mathbf{r}),
     \label{eq:H_rot_frame}
     \end{split}
\end{equation}
where $\mathbf{A} = m\Omega\mathbf{e_z}\times \mathbf{r}$ is the synthetic gauge potential in the symmetric gauge, describing a uniform synthetic magnetic field $\mathbf{B} = B\mathbf{e_z} = 2m\Omega\mathbf{\hat{e}_z}$ corresponding to the Coriolis force. Note that no physical rotation of the trap is required in our proposal, as the synthetic magnetic field appears when describing the system from the rotating frame of reference~\footnote{In protocols with explicit trap rotation, a rotating trap ellipticity is used to transfer angular momentum to the atoms, but it is usually ramped to zero at the end of the protocol and the system is then also described in a rotating frame without any actual rotation.}. The effective potential $V(r)=V_{\text{ext}}(r)-\frac{m\Omega^2}{2}r^2$, with $r=|\mathbf{r}|$, includes the external trapping potential and the centrifugal one, and is assumed to be cylindrically symmetric. Here, $g$ is the two-dimensional contact interaction strength (see Appendix) and $\hat{\Psi}(\mathbf{r})$ is the bosonic field operator for the particles in $\ket{2}$.
Assuming the rotation of the reference frame $\Omega$ to be equal to the trap frequency, the centrifugal potential exactly compensates the harmonic part of the confinement, so that only residual anharmonic terms are present in the rotating frame.

\begin{figure}[h]
    \centering
    \includegraphics[width=\linewidth]{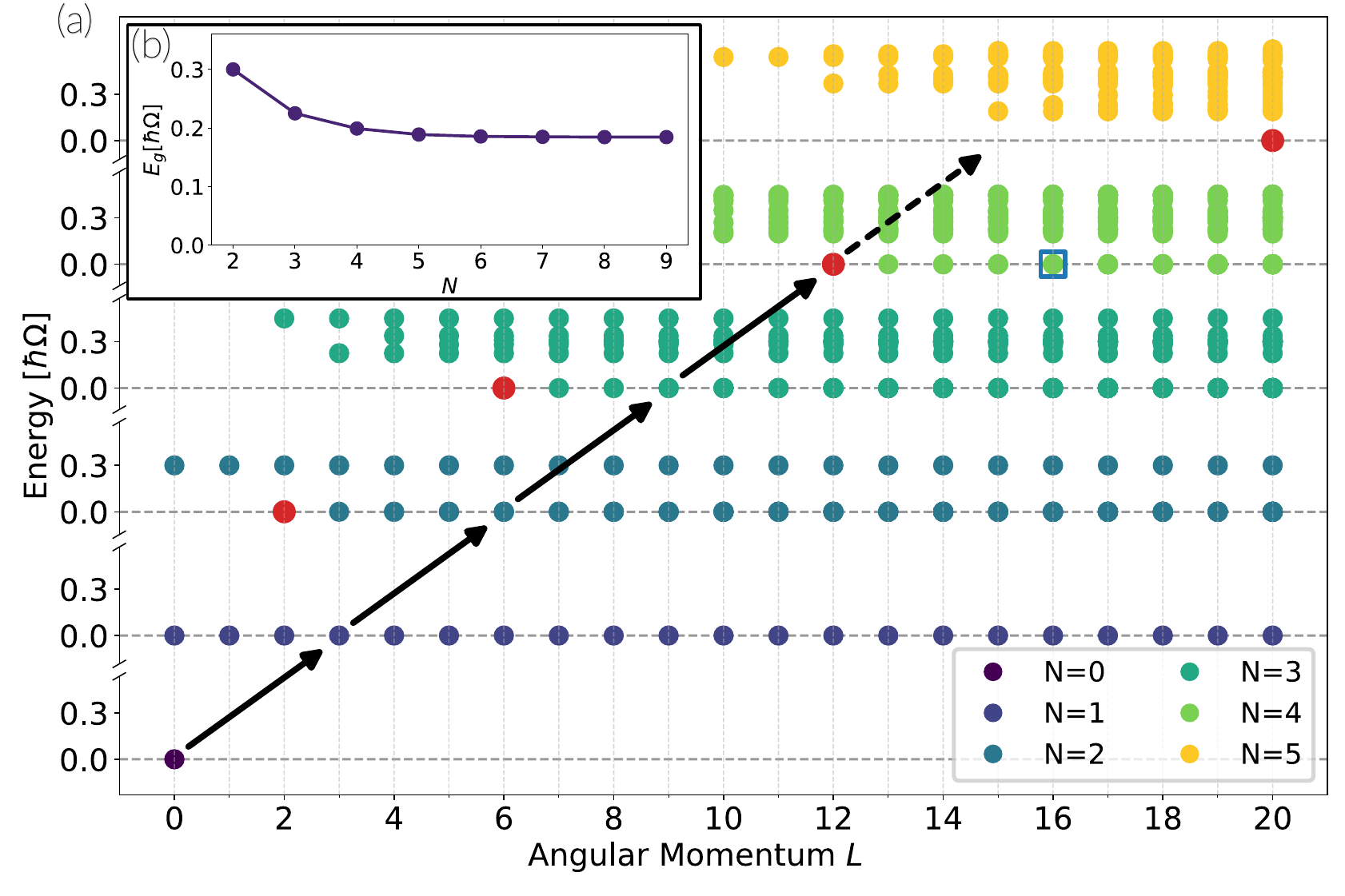}
    \caption{(a) Many-body spectrum of the system in the LLL manifold as a function of the total angular momentum, in the case of no external potential in the rotating frame. Different colors refer to different number sectors. Within each number sector, the energy of the ground state is set to 0. Red dots denote the Laughlin states, of angular momentum $L=N(N-1)$, and the blue square is the state with one quasi-hole and $N=4$. The arrows indicate the transitions coupling the vacuum to the $N=4$ Laughlin state, using a pump with angular momentum $l=3$. Transitions to larger $N$ are suppressed by the many-body gap. (b) Many-body gap above the Laughlin state as a function of $N$ for $g/l_B^2=3.77\hbar\Omega$.}
    \label{fig:spectrum}
\end{figure}

Provided that all energy scales are much smaller than the cyclotron energy $\hbar\omega_c = \hbar B/m = 2\hbar\Omega$, the particles are restricted to live in the lowest Landau level (LLL) manifold of single particle states (see Appendix). In the absence of external potentials, this consists of a macroscopically degenerate subspace of single-particle states labeled by their integer-valued angular momentum quantum number $\alpha\geq 0$, as shown in the $N=1$ sector of \cref{fig:spectrum} (for which $L$ equals $\alpha$).

For a general given $N$, the exact ground state of \cref{eq:H_rot_frame} minimizing angular momentum is the 1/2-Laughlin state~\cite{PhysRevLett.50.1395,cooper2008rapidly}, which lies in the LLL manifold. Within the subspace of given angular momentum quantum number $L = N(N-1)$, the Laughlin state is always a nondegenerate ground state with no interaction energy, separated by a many-body gap $E_g^{(N)}$ proportional to $g$, which tends to a constant as $N\to \infty$~\cite{Regnault2003,JuliaDiaz2012}, as shown in the inset of \cref{fig:spectrum}. The degenerate states with larger angular momenta $L$ correspond to edge and quasi-hole excitations. The degeneracy of these states is lifted if an effective potential $V(r)$ is present in the rotating frame, e.g. a positive quartic potential~\cite{PhysRevA.96.043607}.

The strongly interacting $\ket{2}$ state is coherently coupled with an external electromagnetic field, e.g.\ via a Raman transition, to the BEC in the $\ket{1}$ state, which acts as a reservoir of particles. Assuming that the weakly-interacting BEC is in the mean-field regime and remains in its ground state, the effect of the coherent coupling is described by an additional time-dependent term in the Hamiltonian for the atoms in $\ket{2}$, in the rotating frame, of the form:
\begin{equation}
    \hat{H}_P(t) = \int d^2 r \hbar \Omega_R(\mathbf{r},t)\sqrt{\rho(r)}e^{-i\mu t/\hbar}\hat{\Psi}^\dagger(\mathbf{r})+h.c.
\end{equation}
where $\rho(r)$ is the BEC density, $\mu$ its chemical potential, and $\Omega_R(\mathbf{r},t) = |\Omega_R(r,t)|e^{-il\phi}e^{-i\left(\omega_L' t+\varphi(t)\right)}$ is the two-photon Rabi coupling, and $\omega_L'=\omega_L-l\Omega$ is the frequency difference of the two Raman beams, as seen in the rotating frame.
In order to prepare the desired rotating system state, at least one of the Raman beams is chosen to have a Laguerre-Gaussian profile, so that each particle acquires an orbital angular momentum of $l$ while being transferred from $\ket{1}$ to $\ket{2}$~\cite{Zhang2016,PhysRevLett.133.253401}.
By choosing $\omega_L'+\mu/\hbar = \omega_{1,2}+\Omega$, the residual phase modulation $\varphi(t)$ of the pump can be encoded into an effective detuning $\Delta(t) = \dot{\varphi}(t)$, and the full Hamiltonian in the LLL basis can be recast in the form (see Appendix):
\begin{equation}
\begin{split}
    \hat{H}(t) =& \sum_\alpha V_\alpha \hat{a}_\alpha^\dagger \hat{a}_\alpha + \sum_{\alpha\beta q}G_{\alpha\beta q}\hat{a}^\dagger_\alpha\hat{a}^\dagger_\beta\hat{a}_{\alpha+q}\hat{a}_{\beta-q}  \\
    & - \hbar\Delta(t)\hat{N}+ \hbar F(t)(\hat{a}^\dagger_l+\hat{a}_l).
\end{split}
\label{eq:Ham_pump}
\end{equation}
where $\hat{a}_\alpha^\dagger$ is the creation operator of the LLL of angular momentum $\alpha$, $\hat{N}$ is the operator for the total atom number in the $\ket{2}$ state, $V_\alpha$ encodes the cylindrically symmetric confinement potential $V(r)$, $G_{\alpha\beta q}$ encodes the contact interactions, and the real-valued $F(t)$ is the coupling strength, proportional to the Rabi coupling $\Omega_R$ and the BEC order parameter $\sqrt{\rho}$.
While the focus of this Letter is on rotating clouds, we note that all our results hold for any system described by \cref{eq:Ham_pump}, regardless of how the synthetic magnetic field or the coherent pump are physically implemented, as discussed in the End Matter.
\paragraph{Adiabatic evolution -}
Our idea is to exploit the freedom on the external pump parameters $\Delta(t)$ and $F(t)$, in order to generate a target Laughlin state of $N_t$ particles and angular momentum $L=N_t(N_t-1)$ in the $\ket{2}$ state, starting from the vacuum of $\ket{2}$ atoms. For this purpose, we choose an external pump with single-particle angular momentum $l=N_t-1$, which selects the particle number of the target state. For this specific pump, a chain of states with $N$ particles and angular momentum $L = Nl=N(N_t-1)$ are coupled by the dynamics starting from the empty initial state with $N=L=0$ via a sequence of coherent particle transfers from the reservoir.
As shown in \cref{fig:spectrum}, up to the target particle number $N_t$, all these states have vanishing interaction energy, while for $N\geq N_t+1$ they are shifted up by the many-body gap, since their angular momentum is smaller than the one of the corresponding Laughlin state.

\begin{figure}[h]
    \centering
    \includegraphics[width=\linewidth]{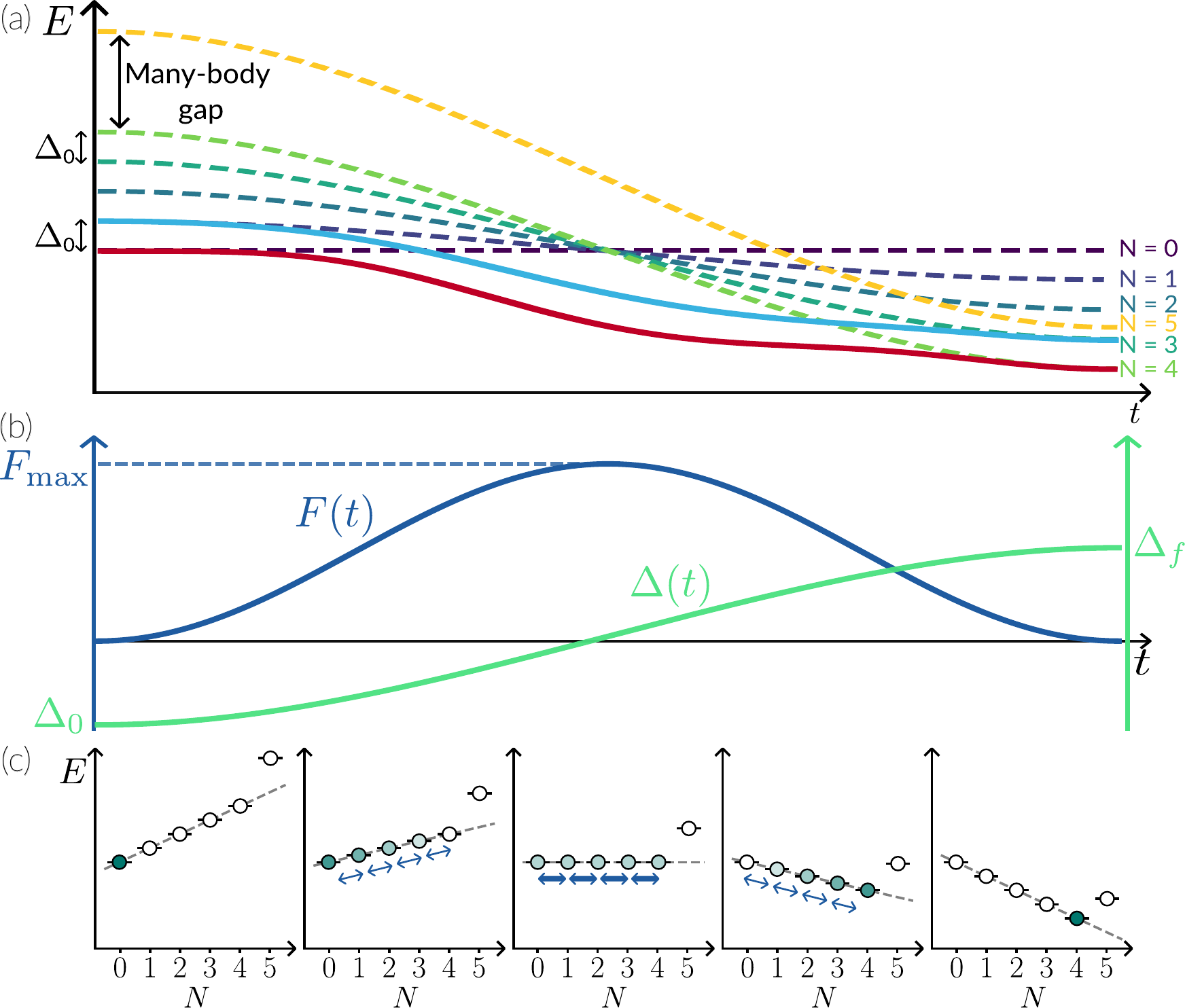}\\
    \caption{(a) Instantaneous eigenstates during the adiabatic evolution with $F=0$ (dashed lines) and first two eigenstates for $F(t)\neq 0$ (solid lines). (b) Sketch of the temporal ramp  of the pump strength $F(t)$ and the detuning $\Delta(t)$. (c) Illustration of the adiabatic scheme as the tunneling in a tilted lattice comprising the states coupled by the pump $F$: initially, the system is in the $N=0$ vacuum state; at the end, it is frozen in the $N_t=4$ Laughlin state protected by the many-body gap (blue dot).}
    \label{fig:adiabatic}
\end{figure}

\cref{fig:adiabatic} shows a pictorial description of the adiabatic evolution for a protocol during which $\Delta(t)$ is swept from negative to positive values as depicted in panel (b).
In panel (a) we show the instantaneous energy levels as a function of time.
The effect of the detuning $\Delta$ is to shift the energy of the different $\hat{N}$ subspaces, acting as a chemical potential. 
At early times, $\Delta<0$, and the lowest energy state is the vacuum. At late times, $\Delta>0$, and the states at finite particle number are pushed down in energy. As long as $0<\Delta<E_g$, the Laughlin state with $N_t$ particles has a lower energy than the gapped $N_t+1$ state. This energy gap blocks population from flowing towards $N\geq N_t+2$ states, which can be neglected (see Appendix). Therefore, the lowest energy state at late times is the target Laughlin state.

The level crossing between states at different $N$ around $\Delta=0$ is avoided by the coherent coupling  of strength $F$. 
This guarantees that a sizable gap is opened by the pump at all times, as illustrated in \cref{fig:adiabatic}(a), and suggests the possibility of  adiabatically transforming the initial vacuum into the target Laughlin state.

\cref{fig:adiabatic}(c) gives a pictorial description of what happens during the adiabatic evolution in terms of a lattice of sites labeled by the particle number $N$ and coupled by the pump which induces nearest-neighbor ''hopping''. 
At $t=0$, the system is prepared in the lowest-energy vacuum state which corresponds to the leftmost site of the lattice. The lattice is then slowly tilted down by changing $\Delta$, while increasing the ''hopping'' $F$, which allows the population to spread along the lattice. At final times, the ''hopping'' $F$ is slowly reduced to freeze the system in the lowest-energy lattice site, which now corresponds to the target Laughlin state.

\begin{figure*}[t]
    \centering
    \includegraphics[width=\linewidth]{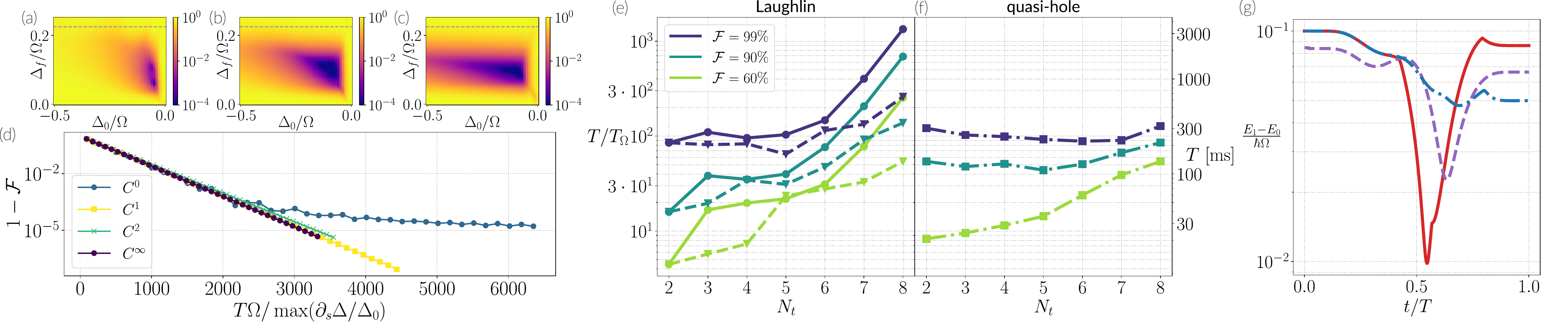}
    \caption{Infidelity with $N_t=4$ Laughlin state for different initial and final detuning, for a $C^\infty$ ramp of ramp time $T = 127T_\Omega$ (318 ms), for (a) $F_{\text{max}}/\Omega=0.05$, (b) $0.1$, (c) $0.2$. The dashed line marks the many-body gap. (d) Infidelity with the $N_t=7$  Laughlin state for ramps with different smoothness, as a function of time, rescaled by the sharpness of $\Delta(s=t/T)$. (e) Ramp time $T$ required to reach a desired fidelity as a function of $N_t$, for a $C^0$ ramp with $\Delta_f = -\Delta_0 = 0.1\Omega$ and $F_{\text{max}}=0.1\Omega$. The solid lines refer to a purely harmonic potential, whereas dashed lines include a quartic potential with $V_0=0.001\hbar \Omega$. The detunings include the energy shift due to the quartic potential. (f) Same plot for quasi-hole state, using a plug potential with $V_0 = 50.0\hbar \Omega$, $\sigma = 0.2l_B$, and a $C^\infty$ ramp. (g) Instantaneous adiabatic gap
    during the evolution to the Laughlin state in the absence (solid red) and presence (dashed purple) of the quartic potential and for the quasi-hole state (dashed-dotted blue line), for the same $C^\infty$ ramp and $N_t=8$: the anharmonic and the plug potentials help keeping a sizable gap also in the middle of the ramp. }
    \label{fig:numerical_results}
\end{figure*}

\paragraph{Numerical results -} In order to put our reasoning on solid quantitative grounds, we have performed numerical simulations of the temporal evolution, \cref{eq:Ham_pump}, under the time-dependent pump for typical parameters of $^7$Li. The effect of losses and experimental imperfections is discussed in the End Matter. We consider ramps of $\Delta(t)$ and $F(t)$ with different degrees of smoothness $C^n$, with $n$ the number of continuous derivatives at the beginning and the end of the protocol.

In \cref{fig:numerical_results} (a,b,c) we show colorplots of the infidelity $1-\mathcal{F} = 1-|\braket{\psi_{\text{La}}}{\psi}|^2$ of the $N_t=4$ final state after a $C^\infty$ smooth ramp of duration $T=127 T_\Omega$, where $T_\Omega = 2\pi/\Omega$ is the trapping period, as a function of the initial and final detunings, $\Delta_{0,f}$. 
The excellent value of the fidelity across a large region in parameter space confirms that our scheme, relying on an adiabatic evolution, is indeed not sensitive to the fine details of the ramp profile, in contrast to, e.g., the $\pi$ pulses of~\cite{PhysRevLett.133.253401}. 

Clear physical explanations can be put forward for the boundaries of the good-fidelity region. 
The initial detuning $\Delta_0$ needs to be negative in order to have the vacuum state as the lowest energy state at the beginning of the protocol. 
Upper bounds on its magnitude $|\Delta_0|$ are 
imposed by the adiabatic condition, as  a larger initial detuning would require a faster change in $\Delta(t)$. Comparison of panels (a,b,c) shows that adiabaticity is restored by a stronger pump which opens a larger gap (see Appendix). Finally, the final detuning $\Delta_f$ must be large enough to well isolate the target state in energy, but not exceed the many-body gap so to keep the Laughlin state as the lowest energy state at the final time.

More details on the dependence of the infidelity on the temporal profile of the ramp are given in \cref{fig:numerical_results}(d).
Here we show the dependence of the infidelity on the ramp duration $T$ for ramps with different smoothness properties.
For ramps at least $C^{1}$, the infidelity shows an exponential decrease with $T$ at a rate set by the sharpness of the ramp profiles. Rescaling time by the derivative of $\Delta(t)$, the curves are found to collapse to the same universal behavior.
Remarkably, the exponential decrease of the infidelity persists up to moderate $T$ also for $C^0$ ramps with a discontinuous first derivative. For longer $T$, this discontinuity is important again and the infidelity recovers the $T^{-2}$ behavior expected from the adiabatic theorem \cite{mackenzie2006perturbative}.
As such, it can actually be advantageous to drop the continuity requirement on the derivative, so  to spread the parameter change throughout the whole evolution and thus reduce the maximum rate of change.

A summary of the achievable fidelity values is given in Fig.~\ref{fig:numerical_results}(e). For moderate-size systems just above the $N_t\leq 3$ experimental state-of-the-art~\cite{leonard2023realization,PhysRevLett.133.253401,doi:10.1126/science.ado3912, kwan2026pfaffianquantumhallstate}, very small infidelities can be obtained already with fast ramps. For instance, a fidelity of $\mathcal{F}=0.99$  can be achieved for $N_t=4$ in $T=95T_\Omega$, which is a 6-fold improvement over theoretical proposals based on a ramp of the trapping frequency and ellipticity~\cite{PhysRevA.70.053612,PhysRevA.103.063325}.
While these fixed-particle-number schemes suffer from the closing of the finite-size-induced gap for growing $N_t$ \cite{PhysRevLett.115.026802,PhysRevB.96.165107, PhysRevB.96.201103}, in our proposal the adiabatic gap is externally controlled by the strength of the coherent coupling and, therefore, can be maintained at a significant size up to large $N_t$. This promising feature is confirmed by our numerical calculations -- restricted to $N_t=8$ only by our computational limitations -- showing that large samples can be realized using ramps of reasonable duration.

On top of a stronger $F_\text{{max}}$ (see End Matter), a further reinforcement of adiabaticity comes from adding an anharmonic confining potential, e.g. $V^{(4)}(r) = V_0r^4/(4l_B^4)$. The improved scaling with $N_t$ shown in \cref{fig:numerical_results}(e) can be explained by the larger adiabatic gap due to broken degeneracy in the $N<N_t$, $L=N(N_t-1)$ sectors by the anharmonic potential, as shown in \cref{fig:numerical_results}(g). At the same time, the anharmonic confinement makes our scheme robust to deconfinement issues. Main limitations of our proposal stem from the imperfect cylindrical symmetry of the trapping potentials and of the Raman beams that tend to slow down the rotation. Also in this respect, the anharmonic potential is beneficial, as it lifts the degeneracy of the harmonic oscillator levels in the laboratory frame, and thus suppresses the slow-down effect.

\paragraph{Quasi-hole state -}
Beyond the Laughlin many-body ground state considered so far, our adiabatic scheme can be extended to excited states including, e.g., quasi-holes. For instance, the state with $N_t$ particles and a single quasi-hole at the center of the cloud has angular momentum $L = N_t^2$ and can be adiabatically generated with a larger angular momentum $l=N_t$ pump.
The degeneracy of the quasi-hole state in the subspace at given $N=N_t$, $L=N_t^2$ can be lifted by means of a spatially localized repulsive potential located at the center, which energetically favors the density depletion of the quasi-hole state. For concreteness, we consider a Gaussian shape for this plug potential $V^{\text{plug}}(r) = V_0e^{-x^2/(2\sigma^2)}/(\sqrt{2\pi\sigma^2})$.

The efficiency of our adiabatic scheme also in this case is showcased by the excellent fidelities shown in \cref{fig:numerical_results} (f)~\footnote{Here, a $C^\infty$ ramp was used as the power-law behavior of the $C^0$ ramp appears in this case already at shorter times.}.  Remarkably, the slower growth of the ramp time required for the generation of quasi-hole states compared to the Laughlin states at larger $N_t$ is due to the larger adiabatic gap opened by the plug potential, as shown in \cref{fig:numerical_results} (g). A combination of suitably designed repulsive plug potentials and sharp potentials on the edges of the FQH cloud~\cite{yao2024observation} can be considered to generate multiply-charged quasi-holes states~\cite{macaluso2020charge}. Moving and reshaping the potential then allows for the manipulation of the quasi-holes and the detection of their fractional charge~\cite{Wang2022} and statistics~\cite{macaluso2020charge,umucalilar2018time,Paredes2001}.

\paragraph{Conclusions -}
In this Letter, we have presented a novel scheme for the preparation of Laughlin states of strongly interacting bosonic atoms. Our proposal is based on an adiabatic passage in a larger Hilbert space allowing the coherent transfer of particles from a Bose-Einstein condensate. This allows to find a path that avoids crossing fixed-particle-number phase transitions. 
As such, the adiabatic gap is not set by finite-size many-body properties, but is tunable via the external drive and the confinement potential: as demonstrated by our numerical simulations up to eight particles, this offers significant improvement in preparation times and is promising in view of the realization of FQH samples far above the current experimental state-of-the-art~\cite{PhysRevLett.133.253401,leonard2023realization}, up to a few tens of particles.

While this Letter focused on atomic systems, our proposal is directly applicable to any system described by the Hamiltonian \cref{eq:Ham_pump}, in particular photonic systems  where tunable coherent pumps are naturally adopted~\cite{Carusotto2013}, synthetic magnetic fields combined with strong interactions are routinely realized \cite{clark2020observation, doi:10.1126/science.ado3912}, and long lifetimes are available for microwave photons~\cite{RevModPhys.93.025005}. Our proposal opens the way to the investigation of the exciting and elusive properties of fractional quantum Hall fluids in synthetic quantum matter.

\paragraph{Acknowledgments}
We acknowledge useful discussion with Lara Giebeler and Maciej Gałka.
The authors acknowledge support by the Deutsche Forschungsgemeinschaft (DFG, German Research Foundation) via the Research Unit FOR 5688 (Project No. 521530974). I.C. acknowledges support by the
Provincia Autonoma di Trento; by the Q@TN Initiative;
by the National Quantum Science and Technology Institute through the PNRR MUR Project under Grant PE0000023-NQSTI, co-funded by the European Union - NextGeneration EU.

\bibliography{references}

@article{Chikkatur2000,
  title = {Suppression and Enhancement of Impurity Scattering in a Bose-Einstein Condensate},
  author = {Chikkatur, A. P. and G\"orlitz, A. and Stamper-Kurn, D. M. and Inouye, S. and Gupta, S. and Ketterle, W.},
  journal = {Phys. Rev. Lett.},
  volume = {85},
  issue = {3},
  pages = {483--486},
  numpages = {0},
  year = {2000},
  month = {Jul},
  publisher = {American Physical Society},
  doi = {10.1103/PhysRevLett.85.483},
  url = {https://link.aps.org/doi/10.1103/PhysRevLett.85.483}
}

@article{Ferrier-Barbut2014,
author = {I. Ferrier-Barbut  and M. Delehaye  and S. Laurent  and A. T. Grier  and M. Pierce  and B. S. Rem  and F. Chevy  and C. Salomon },
title = {A mixture of Bose and Fermi superfluids},
journal = {Science},
volume = {345},
number = {6200},
pages = {1035-1038},
year = {2014},
doi = {10.1126/science.1255380},
URL = {https://www.science.org/doi/abs/10.1126/science.1255380},
eprint = {https://www.science.org/doi/pdf/10.1126/science.1255380}
}

@Article{Wang2022,
	title={{Measurable signatures of bosonic fractional Chern insulator states and their fractional excitations in a quantum-gas microscope}},
	author={Botao Wang and Xiao-Yu Dong and André Eckardt},
	journal={SciPost Phys.},
	volume={12},
	pages={095},
	year={2022},
	publisher={SciPost},
	doi={10.21468/SciPostPhys.12.3.095},
	url={https://scipost.org/10.21468/SciPostPhys.12.3.095},
}

@article{Regnault2003,
  title = {Quantum Hall Fractions in Rotating Bose-Einstein Condensates},
  author = {Regnault, N. and Jolicoeur, Th.},
  journal = {Phys. Rev. Lett.},
  volume = {91},
  issue = {3},
  pages = {030402},
  numpages = {4},
  year = {2003},
  month = {Jul},
  publisher = {American Physical Society},
  doi = {10.1103/PhysRevLett.91.030402},
  url = {https://link.aps.org/doi/10.1103/PhysRevLett.91.030402}
}

@article{JuliaDiaz2012,
  title = {Fractional quantum Hall states of few bosonic atoms in geometric gauge fields},
  author = {Juliá-Díaz, B. and Graß, T. and Barberán, N. and Lewenstein, M.},
  journal = {New Journal of Physics},
  volume = {14},
  number = {5},
  pages = {055003},
  year = {2012},
  publisher = {IOP Publishing},
  doi = {10.1088/1367-2630/14/5/055003},
  url = {https://iopscience.iop.org/article/10.1088/1367-2630/14/5/055003}
}

@article{Fletcher2021,
author = {Richard J. Fletcher  and Airlia Shaffer  and Cedric C. Wilson  and Parth B. Patel  and Zhenjie Yan  and Valentin Crépel  and Biswaroop Mukherjee  and Martin W. Zwierlein },
title = {Geometric squeezing into the lowest Landau level},
journal = {Science},
volume = {372},
number = {6548},
pages = {1318-1322},
year = {2021},
doi = {10.1126/science.aba7202},
URL = {https://www.science.org/doi/abs/10.1126/science.aba7202},
}

@article{Paredes2001,
  title = {$\frac{1}{2}$-Anyons in Small Atomic Bose-Einstein Condensates},
  author = {Paredes, B. and Fedichev, P. and Cirac, J. I. and Zoller, P.},
  journal = {Phys. Rev. Lett.},
  volume = {87},
  issue = {1},
  pages = {010402},
  numpages = {4},
  year = {2001},
  month = {Jun},
  publisher = {American Physical Society},
  doi = {10.1103/PhysRevLett.87.010402},
  url = {https://link.aps.org/doi/10.1103/PhysRevLett.87.010402}
}

@article{Zhang2016,
  title = {Creating fractional quantum Hall states with atomic clusters using light-assisted insertion of angular momentum},
  author = {Zhang, Junyi and Beugnon, J\'er\^ome and Nascimbene, Sylvain},
  journal = {Phys. Rev. A},
  volume = {94},
  issue = {4},
  pages = {043610},
  numpages = {8},
  year = {2016},
  month = {Oct},
  publisher = {American Physical Society},
  doi = {10.1103/PhysRevA.94.043610},
  url = {https://link.aps.org/doi/10.1103/PhysRevA.94.043610}
}

@article{PhysRevLett.50.1395,
  title = {Anomalous Quantum Hall Effect: An Incompressible Quantum Fluid with Fractionally Charged Excitations},
  author = {Laughlin, R. B.},
  journal = {Phys. Rev. Lett.},
  volume = {50},
  issue = {18},
  pages = {1395--1398},
  numpages = {0},
  year = {1983},
  month = {May},
  publisher = {American Physical Society},
  doi = {10.1103/PhysRevLett.50.1395},
  url = {https://link.aps.org/doi/10.1103/PhysRevLett.50.1395}
}

@article{10.1063/1.5131023,
    author = {Hulet, Randall G. and Nguyen, Jason H. V. and Senaratne, Ruwan},
    title = {Methods for preparing quantum gases of lithium},
    journal = {Review of Scientific Instruments},
    volume = {91},
    number = {1},
    pages = {011101},
    year = {2020},
    month = {01},
    issn = {0034-6748},
    doi = {10.1063/1.5131023},
}

@article{PhysRevA.64.012706,
  title = {Interatomic collisions in a tightly confined Bose gas},
  author = {Petrov, D. S. and Shlyapnikov, G. V.},
  journal = {Phys. Rev. A},
  volume = {64},
  issue = {1},
  pages = {012706},
  numpages = {14},
  year = {2001},
  month = {Jun},
  publisher = {American Physical Society},
  doi = {10.1103/PhysRevA.64.012706},
  url = {https://link.aps.org/doi/10.1103/PhysRevA.64.012706}
}

@article{10.1098/rspa.1935.0010,
    author = {Bethe, H. and Peierls, R.},
    title = {Quantum theory of the diplon},
    journal = {Proceedings of the Royal Society of London. A. Mathematical and Physical Sciences},
    volume = {148},
    number = {863},
    pages = {146-156},
    year = {1935},
    month = {01},
    issn = {0080-4630},
    doi = {10.1098/rspa.1935.0010},
}

@article{PhysRevLett.84.2551,
  title = {Bose-Einstein Condensation in Quasi-2D Trapped Gases},
  author = {Petrov, D. S. and Holzmann, M. and Shlyapnikov, G. V.},
  journal = {Phys. Rev. Lett.},
  volume = {84},
  issue = {12},
  pages = {2551--2555},
  numpages = {0},
  year = {2000},
  month = {Mar},
  publisher = {American Physical Society},
  doi = {10.1103/PhysRevLett.84.2551},
  url = {https://link.aps.org/doi/10.1103/PhysRevLett.84.2551}
}

@article{PhysRevLett.45.494,
  title = {New Method for High-Accuracy Determination of the Fine-Structure Constant Based on Quantized Hall Resistance},
  author = {Klitzing, K. v. and Dorda, G. and Pepper, M.},
  journal = {Phys. Rev. Lett.},
  volume = {45},
  issue = {6},
  pages = {494--497},
  numpages = {0},
  year = {1980},
  month = {Aug},
  publisher = {American Physical Society},
  doi = {10.1103/PhysRevLett.45.494},
  url = {https://link.aps.org/doi/10.1103/PhysRevLett.45.494}
}

@article{PhysRevLett.48.1559,
  title = {Two-Dimensional Magnetotransport in the Extreme Quantum Limit},
  author = {Tsui, D. C. and Stormer, H. L. and Gossard, A. C.},
  journal = {Phys. Rev. Lett.},
  volume = {48},
  issue = {22},
  pages = {1559--1562},
  numpages = {0},
  year = {1982},
  month = {May},
  publisher = {American Physical Society},
  doi = {10.1103/PhysRevLett.48.1559},
  url = {https://link.aps.org/doi/10.1103/PhysRevLett.48.1559}
}

@article{RevModPhys.80.1083,
  title = {Non-Abelian anyons and topological quantum computation},
  author = {Nayak, Chetan and Simon, Steven H. and Stern, Ady and Freedman, Michael and Das Sarma, Sankar},
  journal = {Rev. Mod. Phys.},
  volume = {80},
  issue = {3},
  pages = {1083--1159},
  numpages = {0},
  year = {2008},
  month = {Sep},
  publisher = {American Physical Society},
  doi = {10.1103/RevModPhys.80.1083},
  url = {https://link.aps.org/doi/10.1103/RevModPhys.80.1083}
}

@article{clark2020observation,
  title={Observation of Laughlin states made of light},
  author={Clark, Logan W and Schine, Nathan and Baum, Claire and Jia, Ningyuan and Simon, Jonathan},
  journal={Nature},
  volume={582},
  number={7810},
  pages={41--45},
  year={2020},
  publisher={Nature Publishing Group UK London},
  doi={10.1038/s41586-020-2318-5}
}

@article{PhysRevLett.133.253401,
  title = {Realization of a Laughlin State of Two Rapidly Rotating Fermions},
  author = {Lunt, Philipp and Hill, Paul and Reiter, Johannes and Preiss, Philipp M. and Ga\l{}ka, Maciej and Jochim, Selim},
  journal = {Phys. Rev. Lett.},
  volume = {133},
  issue = {25},
  pages = {253401},
  numpages = {7},
  year = {2024},
  month = {Dec},
  publisher = {American Physical Society},
  doi = {10.1103/PhysRevLett.133.253401},
  url = {https://link.aps.org/doi/10.1103/PhysRevLett.133.253401}
}

@article{leonard2023realization,
  title={Realization of a fractional quantum Hall state with ultracold atoms},
  author={L{\'e}onard, Julian and Kim, Sooshin and Kwan, Joyce and Segura, Perrin and Grusdt, Fabian and Repellin, C{\'e}cile and Goldman, Nathan and Greiner, Markus},
  journal={Nature},
  volume={619},
  number={7970},
  pages={495--499},
  year={2023},
  publisher={Nature Publishing Group UK London},
  doi={10.1038/s41586-023-06122-4}
}

@article{PhysRevA.96.043607,
  title = {Hard-wall confinement of a fractional quantum Hall liquid},
  author = {Macaluso, E. and Carusotto, I.},
  journal = {Phys. Rev. A},
  volume = {96},
  issue = {4},
  pages = {043607},
  numpages = {14},
  year = {2017},
  month = {Oct},
  publisher = {American Physical Society},
  doi = {10.1103/PhysRevA.96.043607},
  url = {https://link.aps.org/doi/10.1103/PhysRevA.96.043607}
}

@article{cooper2008rapidly,
  title={Rapidly rotating atomic gases},
  author={Cooper, Nigel R},
  journal={Advances in Physics},
  volume={57},
  number={6},
  pages={539--616},
  year={2008},
  publisher={Taylor \& Francis},
  doi={10.1080/00018730802564122}
}

@article{macaluso2020charge,
  title={Charge and statistics of lattice quasiholes from density measurements: A tree tensor network study},
  author={Macaluso, Elia and Comparin, Tommaso and Umucal{\i}lar, Rifat Onur and Gerster, Matthias and Montangero, Simone and Rizzi, Matteo and Carusotto, Iacopo},
  journal={Physical Review Research},
  volume={2},
  number={1},
  pages={013145},
  year={2020},
  publisher={APS},
  doi={10.1103/PhysRevResearch.2.013145}
}

@article{PhysRevLett.113.155301,
  title = {Topological Growing of Laughlin States in Synthetic Gauge Fields},
  author = {Grusdt, Fabian and Letscher, Fabian and Hafezi, Mohammad and Fleischhauer, Michael},
  journal = {Phys. Rev. Lett.},
  volume = {113},
  issue = {15},
  pages = {155301},
  numpages = {5},
  year = {2014},
  month = {Oct},
  publisher = {American Physical Society},
  doi = {10.1103/PhysRevLett.113.155301}
}

@article{
doi:10.1126/science.ado3912,
author = {Can Wang  and Feng-Ming Liu  and Ming-Cheng Chen  and He Chen  and Xian-He Zhao  and Chong Ying  and Zhong-Xia Shang  and Jian-Wen Wang  and Yong-Heng Huo  and Cheng-Zhi Peng  and Xiaobo Zhu  and Chao-Yang Lu  and Jian-Wei Pan },
title = {Realization of fractional quantum Hall state with interacting photons},
journal = {Science},
volume = {384},
number = {6695},
pages = {579-584},
year = {2024},
doi = {10.1126/science.ado3912}}

@article{Carusotto2013,
  title = {Quantum fluids of light},
  author = {Carusotto, Iacopo and Ciuti, Cristiano},
  journal = {Rev. Mod. Phys.},
  volume = {85},
  issue = {1},
  pages = {299--366},
  numpages = {0},
  year = {2013},
  month = {Feb},
  publisher = {American Physical Society},
  doi = {10.1103/RevModPhys.85.299},
}

@article{RevModPhys.93.025005,
  title = {Circuit quantum electrodynamics},
  author = {Blais, Alexandre and Grimsmo, Arne L. and Girvin, S. M. and Wallraff, Andreas},
  journal = {Rev. Mod. Phys.},
  volume = {93},
  issue = {2},
  pages = {025005},
  numpages = {72},
  year = {2021},
  month = {May},
  publisher = {American Physical Society},
  doi = {10.1103/RevModPhys.93.025005}
}

@article{PhysRevLett.49.405,
  title = {Quantized Hall Conductance in a Two-Dimensional Periodic Potential},
  author = {Thouless, D. J. and Kohmoto, M. and Nightingale, M. P. and den Nijs, M.},
  journal = {Phys. Rev. Lett.},
  volume = {49},
  issue = {6},
  pages = {405--408},
  numpages = {0},
  year = {1982},
  month = {Aug},
  publisher = {American Physical Society},
  doi = {10.1103/PhysRevLett.49.405}
}

@article{PhysRevLett.52.1583,
  title = {Statistics of Quasiparticles and the Hierarchy of Fractional Quantized Hall States},
  author = {Halperin, B. I.},
  journal = {Phys. Rev. Lett.},
  volume = {52},
  issue = {18},
  pages = {1583--1586},
  numpages = {0},
  year = {1984},
  month = {Apr},
  publisher = {American Physical Society},
  doi = {10.1103/PhysRevLett.52.1583}
}

@article{PhysRevLett.53.722,
  title = {Fractional Statistics and the Quantum Hall Effect},
  author = {Arovas, Daniel and Schrieffer, J. R. and Wilczek, Frank},
  journal = {Phys. Rev. Lett.},
  volume = {53},
  issue = {7},
  pages = {722--723},
  numpages = {0},
  year = {1984},
  month = {Aug},
  publisher = {American Physical Society},
  doi = {10.1103/PhysRevLett.53.722}
}

@article{MOORE1991362,
title = {Nonabelions in the fractional quantum hall effect},
journal = {Nuclear Physics B},
volume = {360},
number = {2},
pages = {362-396},
year = {1991},
issn = {0550-3213},
doi = {https://doi.org/10.1016/0550-3213(91)90407-O},
author = {Gregory Moore and Nicholas Read}
}

@article{Wen01101995,
author = {Xiao-Gang Wen},
title = {Topological orders and edge excitations in fractional quantum Hall states},
journal = {Advances in Physics},
volume = {44},
number = {5},
pages = {405--473},
year = {1995},
publisher = {Taylor \& Francis},
doi = {10.1080/00018739500101566}
}

@misc{gemelke2010rotatingfewbodyatomicsystems,
      title={Rotating Few-body Atomic Systems in the Fractional Quantum Hall Regime}, 
      author={Nathan Gemelke and Edina Sarajlic and Steven Chu},
      year={2010},
      eprint={1007.2677},
      archivePrefix={arXiv},
      primaryClass={cond-mat.quant-gas},
      url={https://arxiv.org/abs/1007.2677}, 
}

@article{PRXQuantum.3.030302,
  title = {Experimentally Accessible Scheme for a Fractional Chern Insulator in Rydberg Atoms},
  author = {Weber, S. and Bai, R. and Makki, N. and M\"ogerle, J. and Lahaye, T. and Browaeys, A. and Daghofer, M. and Lang, N. and B\"uchler, H. P.},
  journal = {PRX Quantum},
  volume = {3},
  issue = {3},
  pages = {030302},
  numpages = {11},
  year = {2022},
  month = {Jul},
  publisher = {American Physical Society},
  doi = {10.1103/PRXQuantum.3.030302},
  url = {https://link.aps.org/doi/10.1103/PRXQuantum.3.030302}
}

@article{roncaglia2011rotating,
  title={From rotating atomic rings to quantum Hall states},
  author={Roncaglia, Marco and Rizzi, Matteo and Dalibard, Jean},
  journal={Scientific reports},
  volume={1},
  number={1},
  pages={43},
  year={2011},
  publisher={Nature Publishing Group UK London},
  doi={10.1038/srep00043}
}

@article{PhysRevA.103.063325,
  title = {Preparation of the 1/2 Laughlin state with atoms in a rotating trap},
  author = {Andrade, B\'arbara and Kasper, Valentin and Lewenstein, Maciej and Weitenberg, Christof and Gra\ss{}, Tobias},
  journal = {Phys. Rev. A},
  volume = {103},
  issue = {6},
  pages = {063325},
  numpages = {10},
  year = {2021},
  month = {Jun},
  publisher = {American Physical Society},
  doi = {10.1103/PhysRevA.103.063325}
}

@article{PhysRevLett.108.206809,
  title = {Fractional Quantum Hall States of Photons in an Array of Dissipative Coupled Cavities},
  author = {Umucal\ifmmode \imath \else \i \fi{}lar, R. O. and Carusotto, I.},
  journal = {Phys. Rev. Lett.},
  volume = {108},
  issue = {20},
  pages = {206809},
  numpages = {5},
  year = {2012},
  month = {May},
  publisher = {American Physical Society},
  doi = {10.1103/PhysRevLett.108.206809}
}

@article{PhysRevA.70.053612,
  title = {Adiabatic path to fractional quantum Hall states of a few bosonic atoms},
  author = {Popp, M. and Paredes, B. and Cirac, J. I.},
  journal = {Phys. Rev. A},
  volume = {70},
  issue = {5},
  pages = {053612},
  numpages = {6},
  year = {2004},
  month = {Nov},
  publisher = {American Physical Society},
  doi = {10.1103/PhysRevA.70.053612}
}

@article{moulder2012quantized,
  title={Quantized supercurrent decay in an annular Bose-Einstein condensate},
  author={Moulder, Stuart and Beattie, Scott and Smith, Robert P and Tammuz, Naaman and Hadzibabic, Zoran},
  journal={Physical Review A—Atomic, Molecular, and Optical Physics},
  volume={86},
  number={1},
  pages={013629},
  year={2012},
  publisher={APS},
  doi={10.1103/PhysRevA.86.013629}
}

@article{mackenzie2006perturbative,
  title={Perturbative approach to the adiabatic approximation},
  author={MacKenzie, R and Marcotte, E and Paquette, H},
  journal={Physical Review A—Atomic, Molecular, and Optical Physics},
  volume={73},
  number={4},
  pages={042104},
  year={2006},
  publisher={APS},
  doi={10.1103/PhysRevA.73.042104}
}

@article{fetter1998theory,
  title={Theory of a dilute low-temperature trapped Bose condensate},
  author={Fetter, Alexander L},
  journal={arXiv preprint cond-mat/9811366},
  year={1998},
  doi={10.48550/arXiv.cond-mat/9811366}
}

@article{dutta2018coherent,
  title={Coherent generation of photonic fractional quantum Hall states in a cavity and the search for anyonic quasiparticles},
  author={Dutta, Shovan and Mueller, Erich J},
  journal={Physical Review A},
  volume={97},
  number={3},
  pages={033825},
  year={2018},
  publisher={APS},
  doi={10.1103/PhysRevA.97.033825}
}

@article{horvath2024bose,
  title={Bose-Einstein condensation of non-ground-state caesium atoms},
  author={Horvath, Milena and Dhar, Sudipta and Das, Arpita and Frye, Matthew D and Guo, Yanliang and Hutson, Jeremy M and Landini, Manuele and N{\"a}gerl, Hanns-Christoph},
  journal={Nature Communications},
  volume={15},
  number={1},
  pages={3739},
  year={2024},
  publisher={Nature Publishing Group UK London},
  doi={10.1038/s41467-024-47760-0}
}

@article{tanzi2018feshbach,
  title={Feshbach resonances in potassium Bose-Bose mixtures},
  author={Tanzi, L and Cabrera, CR and Sanz, J and Cheiney, P and Tomza, M and Tarruell, L},
  journal={Physical Review A},
  volume={98},
  number={6},
  pages={062712},
  year={2018},
  publisher={APS},
  doi={10.1103/PhysRevA.98.062712}
}

@book{pitaevskii2016bose,
  title={Bose-Einstein condensation and superfluidity},
  author={Pitaevskii, Lev and Stringari, Sandro},
  volume={164},
  year={2016},
  publisher={Oxford University Press}
}

@article{umucalilar2018time,
  title={Time-of-flight measurements as a possible method to observe anyonic statistics},
  author={Umucal{\i}lar, RO and Macaluso, Elia and Comparin, Tommaso and Carusotto, Iacopo},
  journal={Physical Review Letters},
  volume={120},
  number={23},
  pages={230403},
  year={2018},
  publisher={APS},
  doi={10.1103/PhysRevLett.120.230403}
}

@article{umucalilar2017generation,
  title={Generation and spectroscopic signatures of a fractional quantum Hall liquid of photons in an incoherently pumped optical cavity},
  author={Umucal{\i}lar, RO and Carusotto, I},
  journal={Physical Review A},
  volume={96},
  number={5},
  pages={053808},
  year={2017},
  publisher={APS},
  doi={10.1103/PhysRevA.96.053808}
}

@article{kurilovich2022stabilizing,
  title={Stabilizing the Laughlin state of light: Dynamics of hole fractionalization},
  author={Kurilovich, Pavel and Kurilovich, Vladislav D and Lebreuilly, Jos{\'e} and Girvin, Steven M},
  journal={SciPost Physics},
  volume={13},
  number={5},
  pages={107},
  year={2022},
  doi={10.21468/SciPostPhys.13.5.107}
}

@article{yao2024observation,
  title={Observation of chiral edge transport in a rapidly rotating quantum gas},
  author={Yao, Ruixiao and Chi, Sungjae and Mukherjee, Biswaroop and Shaffer, Airlia and Zwierlein, Martin and Fletcher, Richard J},
  journal={Nature Physics},
  volume={20},
  number={11},
  pages={1726--1731},
  year={2024},
  publisher={Nature Publishing Group UK London},
  doi={10.1038/s41567-024-02617-7}
}

@misc{kwan2026pfaffianquantumhallstate,
      title={A Pfaffian quantum Hall state of ultracold bosons}, 
      author={Joyce Kwan and Perrin Segura and Yanfei Li and Tizian Blatz and Annie Zhi and Brice Bakkali-Hassani and Annabelle Bohrdt and Martin Greiter and Fabian Grusdt and Markus Greiner},
      year={2026},
      eprint={2606.12409},
      archivePrefix={arXiv},
      primaryClass={cond-mat.quant-gas},
      doi={10.48550/arXiv.2606.12409}
}

@article{PhysRevLett.115.026802,
  title = {Continuous Preparation of a Fractional Chern Insulator},
  author = {Barkeshli, M. and Yao, N. Y. and Laumann, C. R.},
  journal = {Phys. Rev. Lett.},
  volume = {115},
  issue = {2},
  pages = {026802},
  numpages = {5},
  year = {2015},
  month = {Jul},
  publisher = {American Physical Society},
  doi = {10.1103/PhysRevLett.115.026802},
  url = {https://link.aps.org/doi/10.1103/PhysRevLett.115.026802}
}

@article{PhysRevB.96.165107,
  title = {Phase transitions and adiabatic preparation of a fractional Chern insulator in a boson cold-atom model},
  author = {Motruk, Johannes and Pollmann, Frank},
  journal = {Phys. Rev. B},
  volume = {96},
  issue = {16},
  pages = {165107},
  numpages = {6},
  year = {2017},
  month = {Oct},
  publisher = {American Physical Society},
  doi = {10.1103/PhysRevB.96.165107},
  url = {https://link.aps.org/doi/10.1103/PhysRevB.96.165107}
}

@article{PhysRevB.96.201103,
  title = {Realizing and adiabatically preparing bosonic integer and fractional quantum Hall states in optical lattices},
  author = {He, Yin-Chen and Grusdt, Fabian and Kaufman, Adam and Greiner, Markus and Vishwanath, Ashvin},
  journal = {Phys. Rev. B},
  volume = {96},
  issue = {20},
  pages = {201103(R)},
  numpages = {6},
  year = {2017},
  month = {Nov},
  publisher = {American Physical Society},
  doi = {10.1103/PhysRevB.96.201103},
  url = {https://link.aps.org/doi/10.1103/PhysRevB.96.201103}
}

@article{PhysRevA.100.063603,
  title = {Measurement of a $^{7}\mathrm{Li}$ tune-out wavelength by phase-patterned atom interferometry},
  author = {Copenhaver, Eric and Cassella, Kayleigh and Berghaus, Robert and M\"uller, Holger},
  journal = {Phys. Rev. A},
  volume = {100},
  issue = {6},
  pages = {063603},
  numpages = {12},
  year = {2019},
  month = {Dec},
  publisher = {American Physical Society},
  doi = {10.1103/PhysRevA.100.063603},
  url = {https://link.aps.org/doi/10.1103/PhysRevA.100.063603}
}

@article{desbuquois2012superfluid,
  title={Superfluid behaviour of a two-dimensional Bose gas},
  author={Desbuquois, R{\'e}mi and Chomaz, Lauriane and Yefsah, Tarik and L{\'e}onard, Julian and Beugnon, J{\'e}r{\^o}me and Weitenberg, Christof and Dalibard, Jean},
  journal={Nature Physics},
  volume={8},
  number={9},
  pages={645--648},
  year={2012},
  publisher={Nature Publishing Group UK London},
  doi = {10.1038/nphys2378}
}

\clearpage

\appendix
\section{Appendix}

\subsection{Experimental parameters}
In order to get concrete numerical estimations, we consider $^7$Li atoms confined in a quasi-two dimensional pancake trap, with strong axial confinement $\omega_z = 2\pi\cdot 10$~kHz and weaker transverse one $\omega_r = 2\pi \cdot 400$~Hz, corresponding to a magnetic length of $l_B = \sqrt{\hbar/(2m\omega_r)}\approx 1.35$~$\mu$m. We consider the $\ket{F=1,m_F=1}=\ket{2}$ hyperfine state as the FQH state, and the dilute BEC in the $\ket{F=1,m_F=0}=\ket{1}$ state as the bath. Exploiting the Feshbach resonance at $B \approx 737.4$~G~\cite{10.1063/1.5131023}, the scattering length of the FQH atoms can be tuned to around $a_2 = 4000 a_0$, the one of the BEC atoms to $a_1 = 11.8a_0$, and the FQH-BEC one to $a_{1,2} = 27.6a_0$. As detailed in the Appendix, this corresponds to a two-dimensional interaction strength $g/l_B^2=3.77\hbar\Omega$ for the atoms in the state $\ket{2}$.

In our $^7$Li proposal, atoms in $\ket{2}$ see the same confinement as $\ket{1}$, but are weakly interacting, and therefore in the dilute regime of Bose-Einstein condensation. Assuming the BEC (and its coherence length) to be much larger than the FQH cloud, residual thermal excitations lie further away from the center of the trap \cite{fetter1998theory,pitaevskii2016bose}, decreasing their contribution to the pump-driven transition to the FQH state. The dominant effect of mutual interactions is, assuming the dilute BEC to be in the mean field regime, to add a repulsive potential on the FQH atoms following the shape of the condensate
\begin{equation}
    \hat{H}_{sb} \approx g_{sb}\rho_{\text{BEC}}\hat{\Psi}^\dagger \hat{\Psi}\,,
\end{equation}
which can be included in the calibration of the effective confinement of the FQH cloud. 

Another consideration is the spontaneous scattering rate from the Raman beams. For $^7$Li atoms, one typically chooses a single-beam detuning of only 5~GHz by selecting a frequency in between the D1 and D2 lines. For the Rabi frequency considered in the main text, the scattering rate is then $R\approx 0.6s^{-1}$, summing over the two transitions. Considering a $C_0$ ramp, where the average Rabi frequency is $2/3$ of its maximum value, and a typical time of 250 ms, for which excellent fidelities are achieved, each atom has a 10\% probability of scattering a photon. For $N_t=8$, this results in a no-scattering probability of around 43\%. Since photon scattering results in a fast recoil and, thus, a prompt loss of the atom from the system, one can post-select images with the desired number of particles. When working with different atoms which, like $^{133}$Cs, display a much larger fine structure splitting, spontaneous scattering is much less important.

The AC Stark shift induced by the Raman beams may create a dipole potential affecting the atoms, of the order of the Rabi coupling frequency. A strategy to circumvent this is to apply the inhomogeneous Laguerre-Gauss beam carrying angular momentum to the transition coupling the weakly interacting level to the intermediate states. In this case, under a local density approximation, the dipole potential felt by the condensate does not result in a sizable  variation of its chemical potential. The FQH system is then coupled to a Gaussian beam only, whose inhomogeneity can be made negligible by choosing a large waist. Alternatively, one could work close to the tune-out wavelength~\cite{PhysRevA.100.063603}, so that the dipole potential is completely canceled out.

An additional process that may be induced by the finite system-bath interaction is the emission of phonons in the BEC by the rotating FQH atoms. As this would result in a loss of angular momentum in the FQH cloud, it should be suppressed, e.g.\ by increasing the BEC density so that its sound velocity exceeds the typical speed of the FQH atoms. In this regime, neglecting acceleration effects, velocities below the Landau criterion prevent point-like defects to generate phonons. The FQH atoms scatter on the BEC as point-like particles regardless of the spatial extension of the FQH wavefunction, such that the Landau criterion applies as observed in related experiments \cite{Chikkatur2000,Ferrier-Barbut2014}. The FQH atoms travel at a speed $v_{\text{FQH}}\approx \sqrt{N}\omega_r l_B \approx 3.4\sqrt{N}$~mm/s, while, assuming $\rho_{\text{BEC}}\approx 10^{14}$~cm$^{-3}$, the resulting speed of sound $c_{\text{BEC}}\approx 8.1$~mm/s is already marginally sufficient to suppress excitations. A reduction of $v_{\text{FQH}}$ can be obtained with smaller confinement frequencies $\omega_r$ along the plane. Another dissipation mechanism is the nucleation of vortex-antivortex pairs, which has been found to be relevant in experiments using stirring via laser beams \cite{desbuquois2012superfluid}. However, this mechanism becomes only relevant when the moving obstacle becomes comparable to the healing length of the BEC, which is not the case for the point-like scattering between system and BEC.

A further way to limit phonon emission or vortex-pair excitation into the BEC is to slightly shift the spatial position of the BEC and FQH clouds along the $z$ direction, so as to decrease their overlap and suppress their collisional interaction, while still keeping a sizable Raman coupling. Although $^7$Li in the Paschen-Back regime in the high-field seeking states considered here does not allow for spin-dependent potentials, a combination of a high-field seeking and a low-field seeking state would allow splitting the two clouds via a magnetic field gradient. However, not all scattering lengths involved in this case are known. Alternatively, one could consider other atoms with two long-lived states and tunable interactions, like $^{133}$Cs~\cite{horvath2024bose} or $^{41}$K~\cite{tanzi2018feshbach}.
Since the spatial coincidence of two or more particles is strongly suppressed in the Laughlin state, three body or dipolar relaxation, which usually limit the lifetime of strongly interacting atoms, are significantly suppressed. Going beyond the present protocol, such loss terms could even be used to stabilize desired states.

We assume to couple the two atomic levels with two Raman beams, one of which is chosen as a Laguerre-Gaussian beam~\cite{moulder2012quantized}, in order to transfer angular momentum $l$ to the FQH atoms. The strength of the coupling $F$ entering the LLL Hamiltonian \cref{eq:Ham_pump} is determined by an overlap integral between the space-dependent Rabi frequency of the two-photon Raman pump and the single-particle LLL with angular momentum $l$, as detailed in the Appendix. Given that the radial profile of the Laguerre-Gaussian mode and LLL wavefunction are the same, the waist of the Laguerre-Gaussian beam can be chosen such that the intensity of the coupling $F$ does not decrease with $l$. Furthermore, the coupling intensity in enhanced by the macroscopic population of the BEC.

Anisotropies in the trapping potential couple to states with smaller angular momentum, however, they can be controlled to extremely high precision, with a control on the ellipticity $\delta \omega/\omega$ of around $2 \times 10^{-3}$ in TOP traps \cite{Fletcher2021} and to the level of $4 \times 10^{-4}$ in optical traps \cite{PhysRevLett.133.253401}.
The effect of residual anisotropies, as well as other experimental imperfections, is reduced by an anharmonicity in the trapping potential which lifts the degeneracy of the level of the two-dimensional harmonic oscillator~\cite{PhysRevLett.133.253401} and therefore suppresses transitions between states with different angular momentum. Given the additional improvement in preparation fidelity in the presence of a quartic potential, as shown in the main text, we conclude that anharmonicity is extremely advantageous in our scheme.

As a further experimental imperfection, a spatially displaced Laguerre-Gaussian beam would generate field components with $l'<l$, coupling to different states in the spectrum. Given the existence of the many-body gap at $N_t$, these states can only couple to excited states with smaller $N<N_t$. For the final value of the detuning, such states have anyway a higher energy than the target Laughlin state at $N_t$ and therefore remain empty at the end of the adiabatic ramp. Beyond the LLL approximation, the displaced rotation of the Laguerre-Gaussian beam as seen from the rotating frame may induce a slowing down of the sample. This effect is however attenuated in the presence of the anharmonic potential.

\subsection{Effect of larger pump power on adiabaticity}

\begin{figure}[t]
    \centering
    \includegraphics[width=0.7\linewidth]{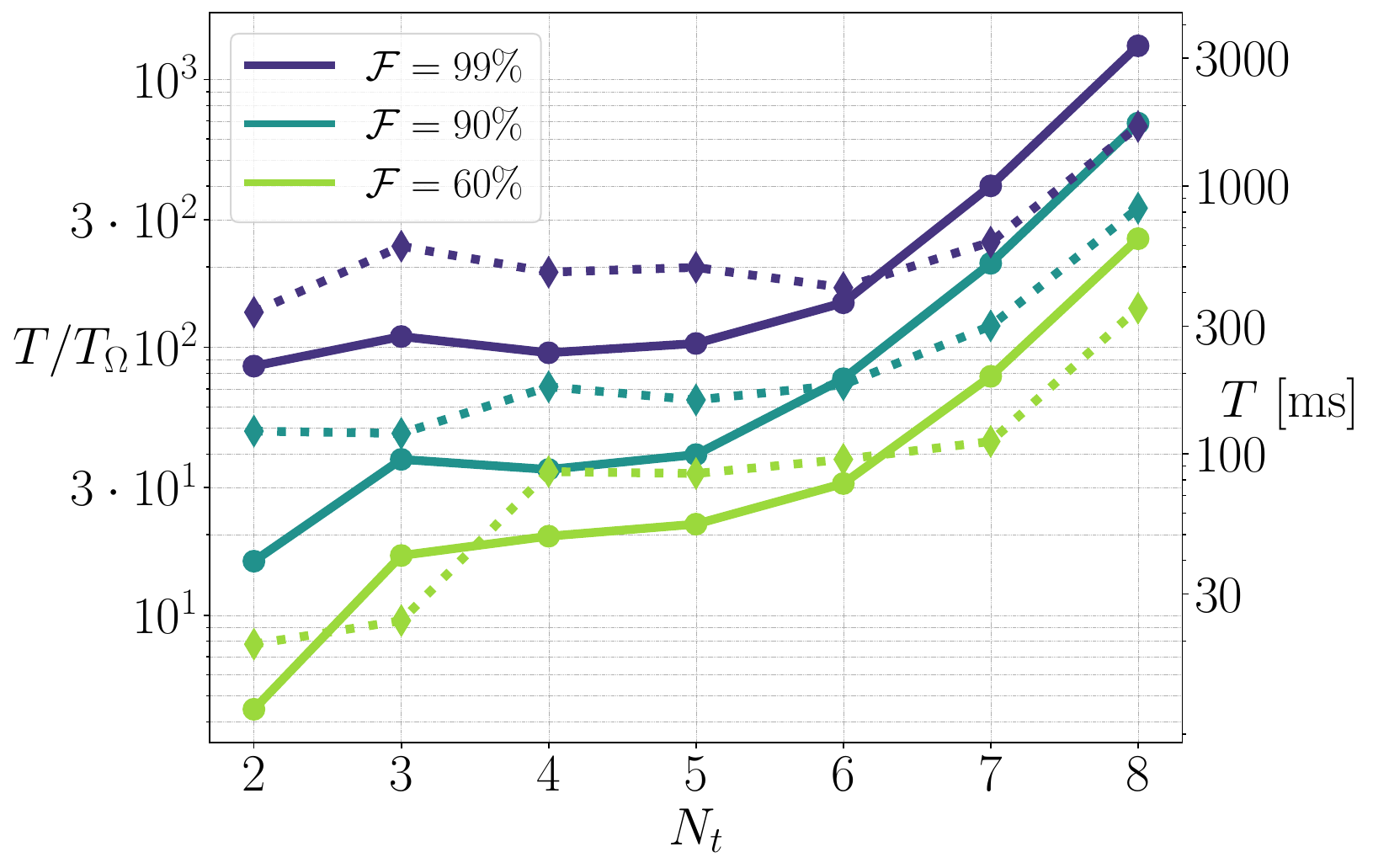}
    \caption{Time required to reach a desired fidelity as a function of $N_t$, for a $C^0$ ramp with $\Delta_f = -\Delta_0 = 0.1\Omega$ and $F_{\text{max}}=0.1\Omega$ (solid lines) or $F_{\text{max}}=0.2\Omega$ (dashed lines).}
    \label{fig:larger_F}
\end{figure}

Given that a larger adiabatic gap opens when $F_{\text{max}}$ increases, a larger $F_{\text{max}}$ can be used in order to better protect the adiabatic dynamics. Although a larger $F_{\text{max}}$ also means a faster variation of $F(t)$, the required timescale for adiabatic dynamics can be reduced for large $N_t$, as shown in \cref{fig:larger_F}.

\FloatBarrier

\subsection{Extension of the scheme to other platforms}
The protocol presented in this Letter is applicable to any system described by \cref{eq:Ham_pump}, i.e. any system with a LLL manifold and a coherent pump which is selective in angular momentum, regardless of the specific implementation of the synthetic gauge field. If the synthetic gauge field is realized in the laboratory frame, then the only modification to the presented scheme is that the laser frequency is not shifted by $l\Omega$.

In addition, even though our scheme relies on rotational symmetry in order to exploit angular momentum conservation, and is therefore naturally described by a static vector potential in the symmetric gauge, it can be extended to any other gauge choice.
In general, gauge invariance imposes that if the vector potential changes as
$\mathbf{A}'(\mathbf{r})=\mathbf{A}(\mathbf{r})+\nabla \alpha (\mathbf{r})$,
then the wavefunction, assuming a charge $q=1$, must acquire a phase given by $\psi'(\mathbf{r},t)=e^{i\alpha(\mathbf{r})/\hbar}\psi(\mathbf{r},t)$.

As an example, if the synthetic magnetic field is realized in the Landau gauge $\mathbf{A}'(\mathbf{r})=B(0,x)$, one gets $\alpha(\mathbf{r}) = \frac{B}{2}xy$. This phase factor must be imprinted on the coherent pump on top of the angular momentum Laguerre-Gauss phase profile. In experiments, this is readily done using spatial light modulators.

\subsection{Determination of the  interaction strength}
\begin{figure}[h]
    \centering
    \includegraphics[width=0.9\linewidth]{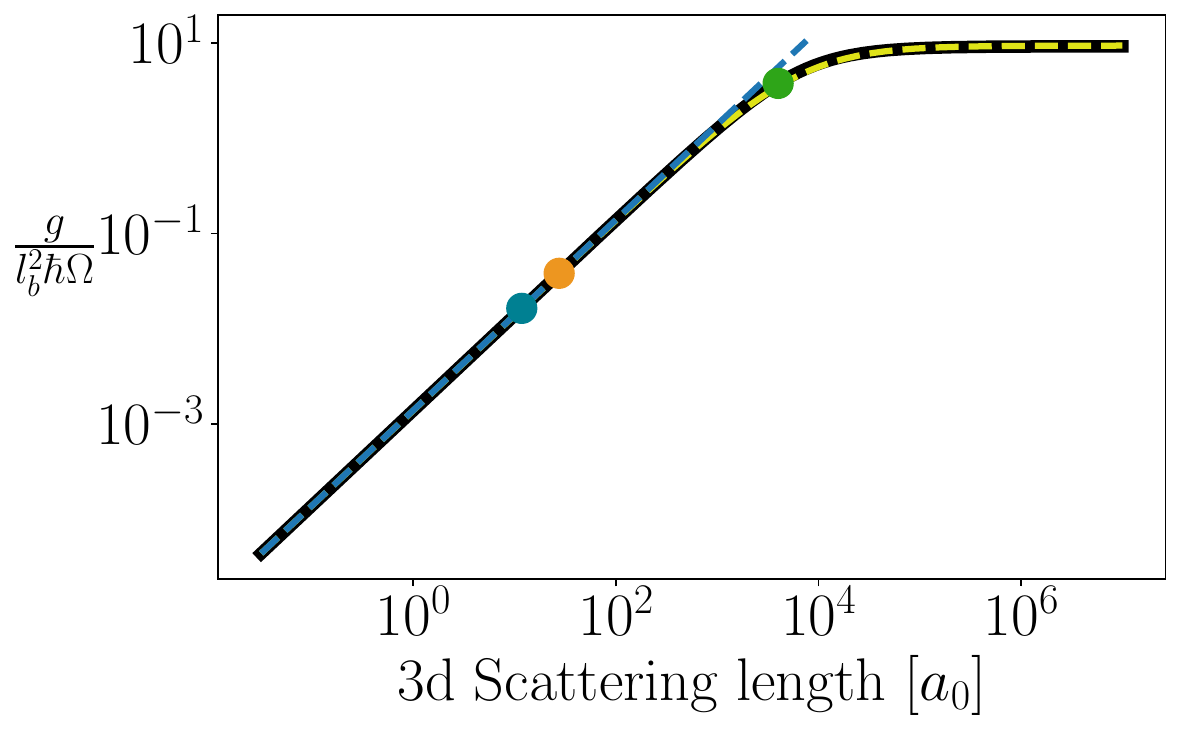}
    \caption{Two-dimensional contact interaction $g$ as a function of the three-dimensional scattering length $a_s$. The solid curve is the result of the numerical analysis explained in the main text, the blue dashed line is the mean-field prediction, whereas the yellow dashed line corresponds to \cref{eq:log_correction} with $ql_z \approx 0.022$. The dots correspond to the values of the interactions between atoms in $\ket{2}$ (green), atoms in $\ket{1}$ (blue), and the mutual interaction (orange).}
    \label{fig:estimate_g_Li}
\end{figure}

In order to estimate the interaction strength $g$ in the 2d FQH cloud, we relate the 3d scattering length to the in-plane one in the quasi-2d regime as~\cite{PhysRevA.64.012706}
\begin{equation}
    a_{2d} \approx 2.08 l_z \exp\left[-\sqrt{\frac{\pi}{2}}\frac{l_z}{a_{s}}\right].
\end{equation}
We then compute the excitation energy of two particles in an harmonic potential imposing the two-dimensional version of the Bethe-Peierls boundary conditions~\cite{10.1098/rspa.1935.0010} $\psi\sim \ln \left(r/a_{2d}\right)$ for $r\to 0$, for the relative motion. Comparing this interaction energy to the $N=2$ many-body gap in the LLL basis, we get an estimate for the contact interaction $g$. 
The result as a function of the 3d scattering length is plotted in \cref{fig:estimate_g_Li}. 

For small scattering length, the expected linear mean-field behaviour is recovered, whereas a logarithmic correction arises for stronger interactions, which fits well with the estimate of~\cite{PhysRevLett.84.2551}:
\begin{equation}
    g = \frac{\sqrt{8\pi}\hbar^2}{m}\frac{1}{l_z/a_s+1/(\sqrt{2\pi})\ln\left(1/(\pi q^2l_z^2)\right)}
    \label{eq:log_correction}
\end{equation}
with $q\ll 1/l_z$ setting the typical scattering energy scale.

From this analysis we find that, for $a_s = 4000a_0$, one gets $g/l_B^2\approx 3.77 \hbar \Omega$, which is the value used in every figure of this work. 

\subsection{Hamiltonian in the LLL basis}
When all energy scales are much smaller than the cyclotron energy $\hbar \omega_c = \hbar B/m = 2\hbar \Omega$, then the FQH system is restricted to live in the lowest Landau level (LLL) basis:
\begin{equation}
    \psi_{\alpha}^{\text{LLL}}(\mathbf{r}) = \frac{1}{l_B\sqrt{2\pi\alpha!}}\left(\frac{re^{-i\phi}}{\sqrt{2}l_B}\right)^\alpha e^{-\frac{r^2}{4l_B^2}}
    \label{eq:LLL}
\end{equation}
which is a macroscopically degenerate subspace labelled by its angular momentum eigenvalue $\alpha \geq 0$. Here, $r$ is the two-dimensional radial coordinate, $\phi$ the azimuthal angle, and $l_B= \sqrt{\hbar/(2m\omega_r)}$ is the magnetic length.
The Hamiltonian describing the FQH system can be written in the LLL manifold, also including an effective central potential in the rotating frame, via the change of variable
\begin{equation}
    \hat{\Psi}(\mathbf{r}) = \sum_\alpha \psi_\alpha^{\text{LLL}}(\mathbf{r})\hat{a}_\alpha
\end{equation}
and reads
\begin{equation}
\begin{split}
    \hat{H}(t) =& \sum_\alpha V_\alpha \hat{a}_\alpha^\dagger\hat{a}_\alpha + \sum_{\alpha\beta q}G_{\alpha\beta q}\hat{a}^\dagger_\alpha\hat{a}^\dagger_\beta\hat{a}_{\alpha+q}\hat{a}_{\beta-q} +  \\
    & - \hbar\Delta(t)\hat{N}+ \hbar F(t)(\hat{a}^\dagger_l+\hat{a}_l).
\end{split}
\end{equation}
where
\begin{equation}
    V_\alpha = \frac{1}{l_B^{2\alpha+2}2^\alpha\alpha!}\int_0^\infty dr r^{2\alpha+1}V(r)e^{-\frac{r^2}{2l_B^2}},
\end{equation}

\begin{equation}
    G_{\alpha\beta q} = \frac{g}{2\pi l_B^2}\frac{(\alpha+\beta)!}{2^{\alpha+\beta+2}\sqrt{\alpha!\beta!(\alpha+q)!(\beta-q)!}},
\end{equation}
and
\begin{equation}
    F = \frac{2\pi}{l_B^{l+1}2^{l/2}\sqrt{2\pi l!}}\int_0^\infty dr r^{l+1}|\Omega_R(r,t)|\sqrt{\rho(r)}e^{-\frac{r^2}{4l_B^2}}.
\end{equation}
In particular, for a quartic potential $V^{(4)}(r) = V_0r^4/(4l_B^4)$
\begin{equation}
    V^{(4)}_\alpha = V_0 (\alpha+1)(\alpha+2).
    \label{eq:quartic}
\end{equation}
For the plug potential $V^{\text{plug}}(r) = \frac{V_0}{\sqrt{2\pi\sigma^2}}e^{-x^2/(2\sigma^2)}$ considered for the quasi-hole generation, one gets
\begin{equation}
    V_\alpha = \frac{V_0}{\sqrt{2\pi\sigma^2}}\left(\frac{\sigma^2}{\sigma^2+l_B^2}\right)^{\alpha+1}.
    \label{eq:plug}
\end{equation}
In the presence of contact interactions, the Laughlin state, which lies in the LLL manifold, is the exact ground state~\cite{PhysRevLett.50.1395,cooper2008rapidly}, and its wavefunction reads 
\begin{equation}
    \psi_\text{La}(\{z_i\},\{\bar{z}_i\}) \sim \prod_{i<j}^N(z_i-z_j)^2e^{-\frac{|z_i|^2}{4l_B^2}}
\end{equation}
where $l_B = \sqrt{\hbar/(2m\Omega)}$ is the magnetic length and $z=re^{i\phi}$, with azimuthal coordinate angle $\phi$.  The many-body gap resulting from the interaction strength used in the main text is around 20\% of the cyclotron energy. Therefore, low-excited states are also well captured by the LLL approximation. Moreover, since the initial (vacuum) and final (Laughlin) state are in the LLL manifold, the contribution of higher Landau levels only modifies the details of the adiabatic dynamics, but not the final target state.

\subsection{Numerical method}
\begin{figure}[h]
    \centering
    \includegraphics[width=0.9\linewidth]{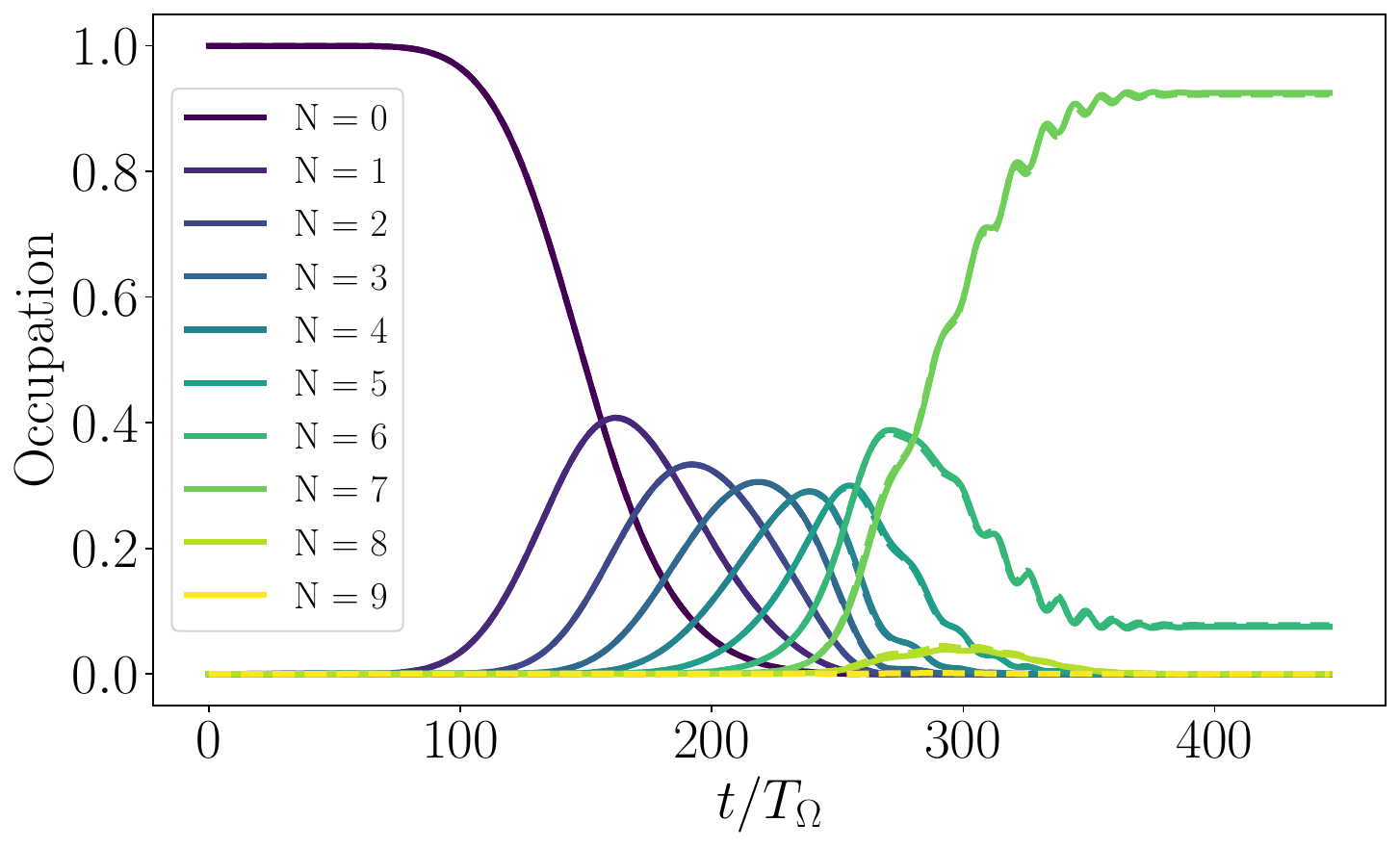}
    \caption{Occupation of the number sectors for a ramp with $T = 446 T_\Omega$ targeting a $N_t=7$ Laughlin state, with the same parameters as in the main text. The truncation of Hilbert space was set to $N_{\text{max}} = 8$ (solid lines) and $N_{\text{max}} = 9$ (dashed lines). The final infidelity is $1-\mathcal{F}\approx 7.484\cdot 10^{-2}$ and $1-\mathcal{F}\approx 7.773\cdot 10^{-2}$ respectively, whereas the maximum occupation of the $N=9$ sector is $1.41\cdot 10^{-3}$.}
    \label{fig:N_max_compare}
\end{figure}

Exact diagonalization in the LLL basis was employed, restricting to the number and angular momentum sectors actually coupled by the considered pump. For the scheme presented in the main text, this means that only the sectors with $N$ particles and angular momentum $L=N(N_t-1)$ were considered, and the Hilbert space was truncated at the $N_t+1$ sector. The negligible effect of the truncation is shown in \cref{fig:N_max_compare}, where we compared with a simulation truncated at a larger value $N_t+2$. The size of the Hilbert space for $N_t=7$, truncated at $N_{\text{max}}=9$, is equal to 107,415, whereas for $N_{\text{max}}=8$ it decreases to 36,053, implying a significant reduction in computational resources. The dimension of the largest Hilbert space considered for the dynamics with $N_t=8$ is 256,249.

\subsection{Ramp shapes}
\begin{figure}[h]
    \centering
    \includegraphics[width=0.9\linewidth]{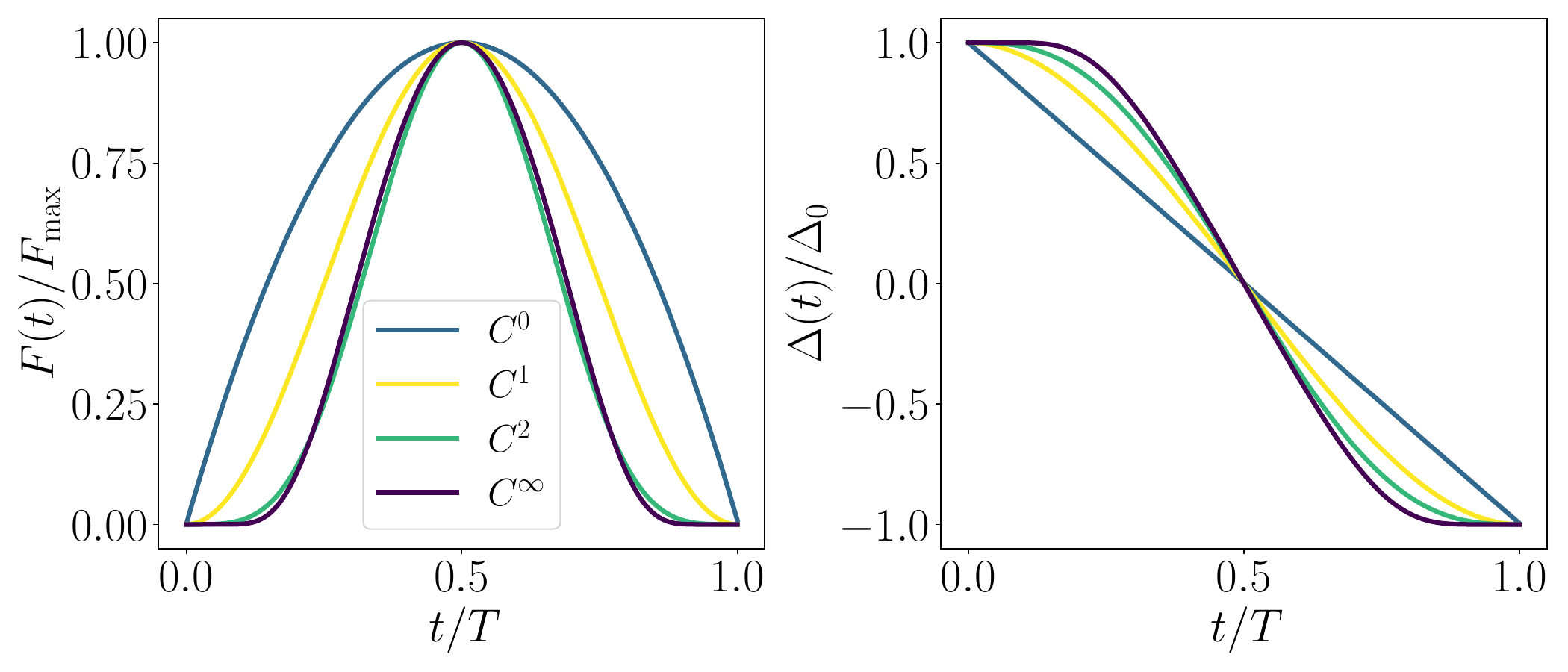}
    \caption{Plot of the temporal profile of the pump strength (left) and of the pump detuning (right), for the four different ramps with different degrees of smoothness used in the text.}
    \label{fig:ramp_profiles}
\end{figure}

The ramp profiles used in the main text are, as a function of $s=t/T$:
\begin{equation}
    F(s) = F_{\text{max}}\begin{cases}
        \exp\left[4-\frac{1}{s(1-s)} \right] & (a) \\
        \sin\left(\pi s\right)^4 & (b)\\
        \sin\left(\pi s\right)^2 & (c)\\
        4s(1-s) & (d)
    \end{cases}
\end{equation}

\begin{equation}
    \Delta(s) = \begin{cases}
        (\Delta_f - \Delta_0) \frac{e^{-1/s}}{e^{-1/s}+e^{-1/(1-s)}} + \Delta_0 & (a) \\
        (\Delta_f-\Delta_0)\left(6s^5-15 s^4+10 s^3\right) +\Delta_0 & (b)\\
        2(\Delta_0-\Delta_f)\left(s^3-\frac{3}{2}s^2\right) +\Delta_0 & (c) \\
        (\Delta_f - \Delta_0) s + \Delta_0 & (d)
    \end{cases}
\end{equation}
corresponding to a smooth $C^\infty$ ramp (a), one with continuous second derivative, $C^2$ (b), one with continuous first derivative, $C^1$ (c), and one which is only continuous $C^0$ (d). Their profile is plotted in \cref{fig:ramp_profiles}.

\subsection{Instantaneous adiabatic gap}
\begin{figure}[t]
    \centering
    \includegraphics[width=0.75\linewidth]{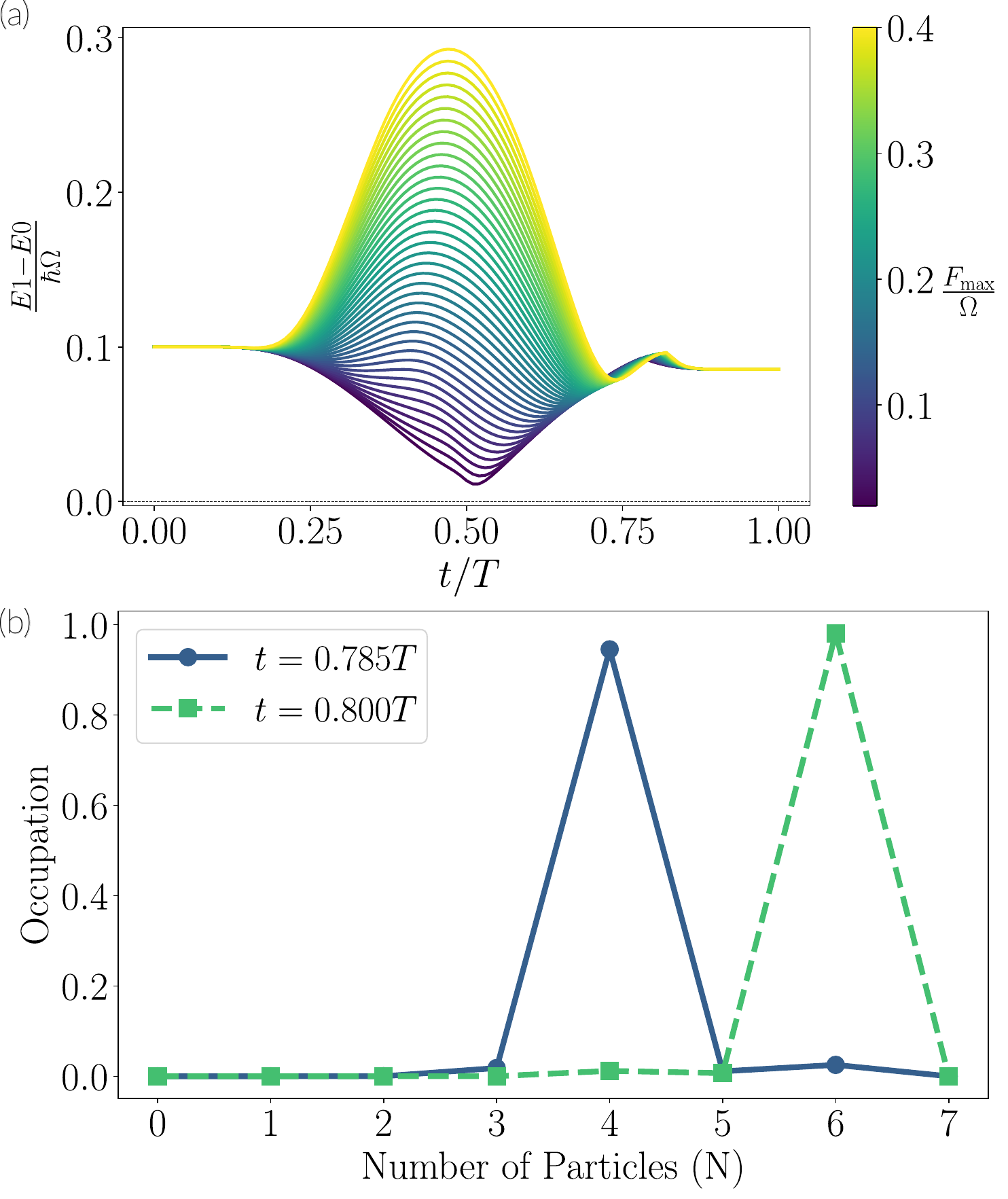}
    \caption{(a) Instantaneous gap for $N_t=5$, for a $C^\infty$ ramp with initial detuning $\Delta_0 = -0.1\Omega$, final detuning $\Delta_f = 0.1\Omega$, for different values of $F_{\text{max}}\in[0.02,0.4]\Omega$. Around $t=0.8T$, the first excited state goes through a level crossing, with its occupation (b) going from the $N_t-1$ to the $N_t+1$ sector.}
    \label{fig:gap_F}
\end{figure}

\cref{fig:gap_F} (a) shows the instantaneous many-body gap in the sectors with $L=N(N_t-1)$, actually coupled by the pump, for different values of the maximum coupling $F_{\text{max}}$ during the $C^\infty$ ramp, which remains finite throughout the whole evolution.
It is interesting to note that excited states are not necessarily gapped, since the one-particle pump may not couple them. Namely, when the energy of the $N_t-1$ sector approaches that of the gapped $N_t+1$, there is a level crossing. The first excited state then transitions from $N_t-1$ to $N_t+1$, as shown in \cref{fig:gap_F} (b), resulting in a cusp of the gap curve.

\subsection{Sequential ramp}
\begin{figure}[b]
    \centering
    \includegraphics[width=0.9\linewidth]{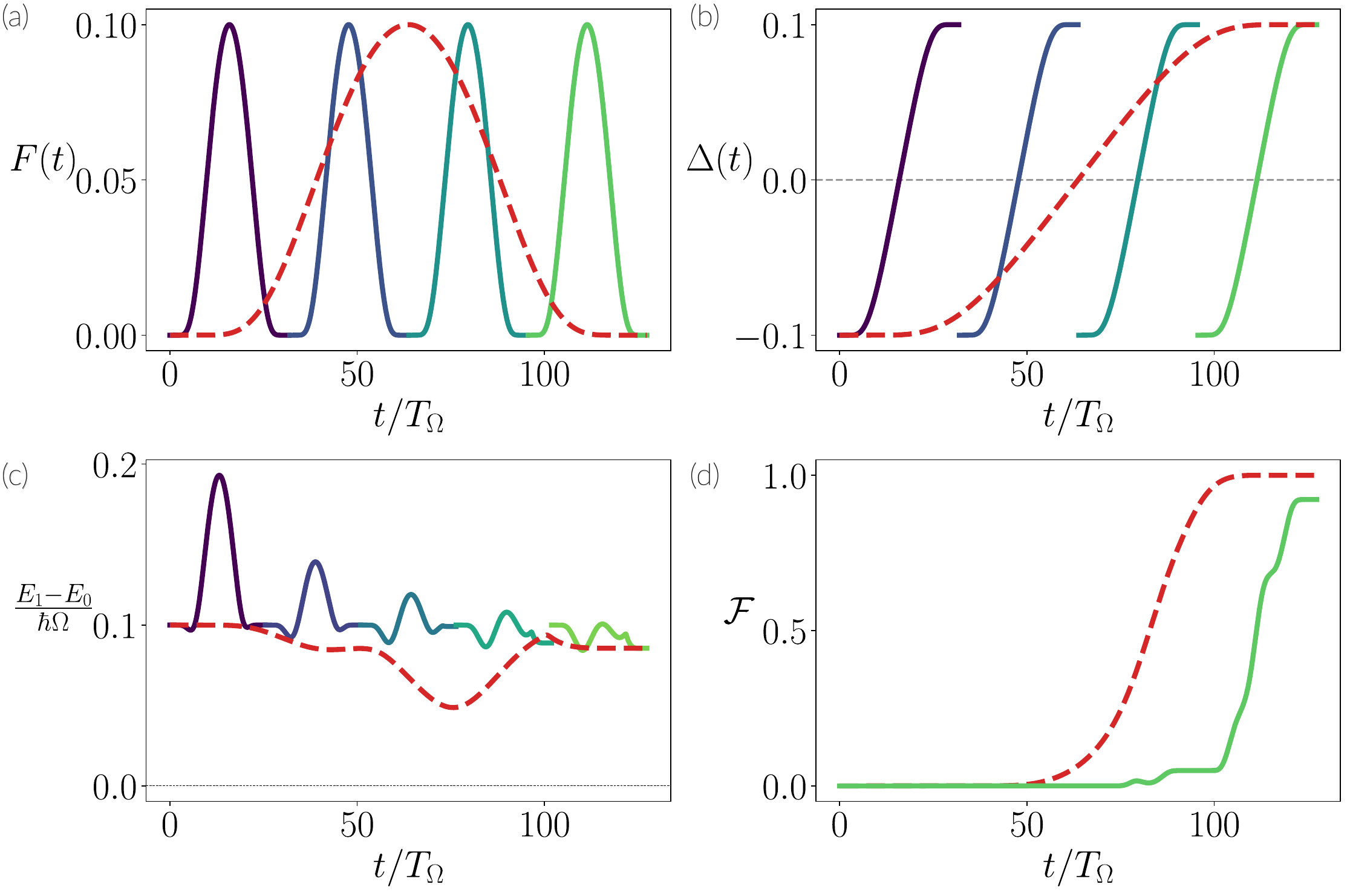}
    \caption{Comparison of the two schemes to target the $N_t=4$ Laughlin state with same ramp time $T = 127T_\Omega$. (a)-(b) Ramp profile for the pump strength (a) and detuning (b) for the one-shot (red dashed) and sequential (solid) schemes. (c) Energy gap during the adiabatic evolution for the one-shot (red dashed) and sequential (solid) schemes. (d) Fidelity with the target Laughlin state during the evolution for the one-shot (red dashed) and sequential (solid) schemes. The one-shot scheme performs much better, with a final infidelity of $1-\mathcal{F} \approx 1.58\cdot 10^{-5}$, compared to the sequential scheme with $1-\mathcal{F}\approx 7.8\cdot 10^{-2}$.}
    \label{fig:compare_schemes}
\end{figure}

\begin{figure}[h]
    \centering
    \includegraphics[width=0.9\linewidth]{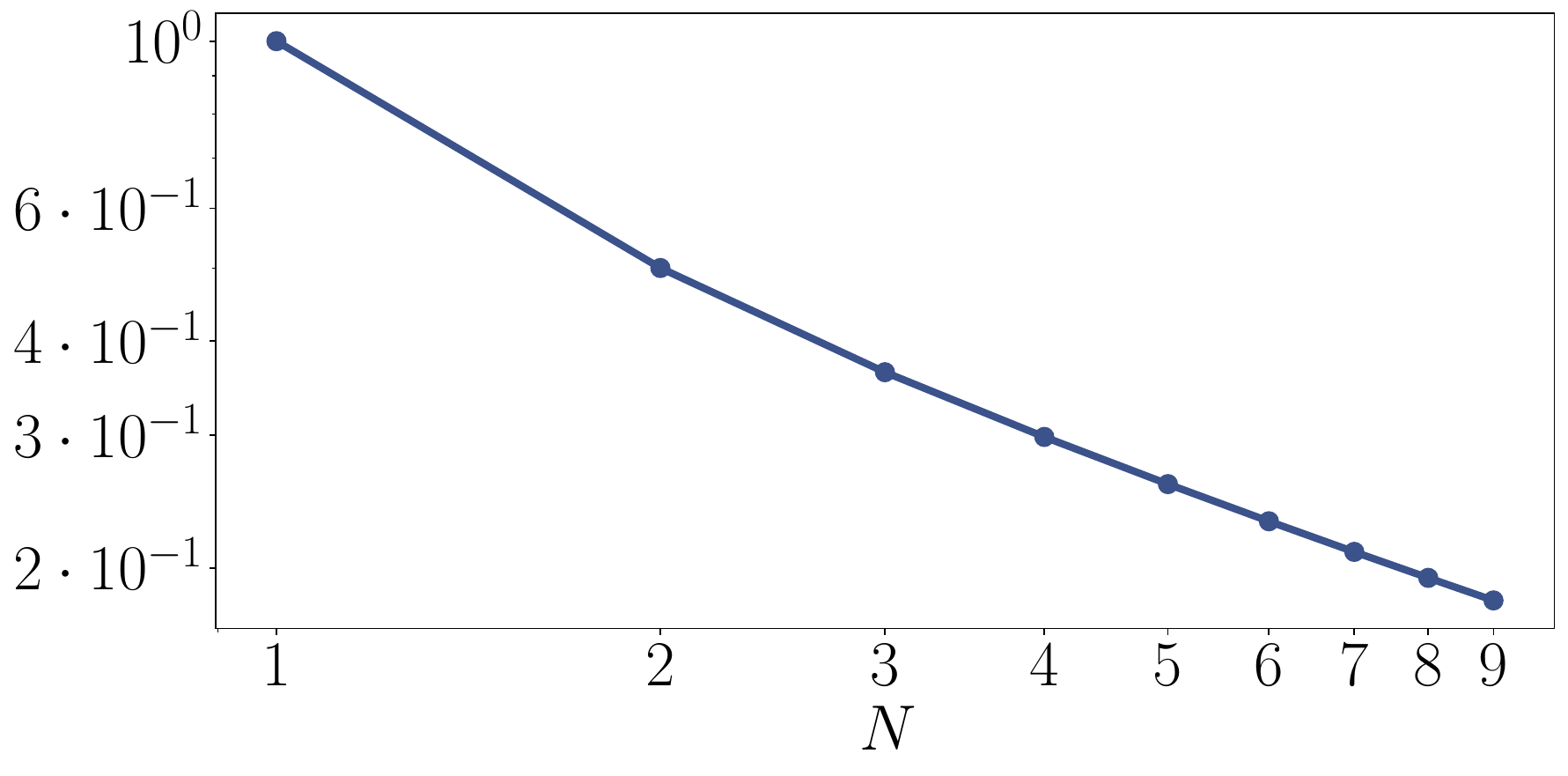}
    \caption{Overlap between a Laughlin state with $N-1$ particles, to which one particle is added with angular momentum $l = 2(N-1)$, and the Laughlin state with $N$ particles. This matrix element, $|\langle \psi_{N}^{\text{La}} | \hat{a}^\dagger_{2(N-1)} | \psi_{N-1}^{\text{La}} \rangle|^2$, sets the magnitude of the gap open by the external coupling.}
    \label{fig:matrix_el_sequential}
\end{figure}

In addition to the one-shot scheme proposed in the main text, we also investigated employing a sequential scheme analogous to \cite{dutta2018coherent}, consisting of $N_t$ individual subramps of $F(t)$ and $\Delta(t)$, as shown in panels (a) and (b) of \cref{fig:compare_schemes}. Within each subramp a pump with angular momentum of $l=2(N-1)$ is used to convert a Laughlin state at $N-1$ particles to the next Laughlin state at $N$ particles. Although this might be the most natural solution, we found that the effect of concatenating $N_t$ different subramps makes the infidelity larger, as shown in \cref{fig:compare_schemes} (d), since any loss in fidelity at the end of a single subramp cannot be recovered and will be rather amplified by the next subramp. 
\cref{fig:compare_schemes} (c) displays the instantaneous gap during the evolution in the two cases: while the gap has a comparable or even larger value in the sequential scheme, this advantage is overcompensated by the effective speed of the variation of the Hamiltonian which is $N_t$ times larger in the sequential scheme and by the need for concatenating $N_t$ subramps, which accumulates errors. Still, a remarkable advantage of the sequential scheme is that the matrix elements of the coupling of one Laughlin state with the next only decay algebraically with $N$, as shown in \cref{fig:matrix_el_sequential}.

\FloatBarrier

\subsection{Post-selected fidelity}
\begin{figure}[h]
    \centering
    \includegraphics[width=\linewidth]{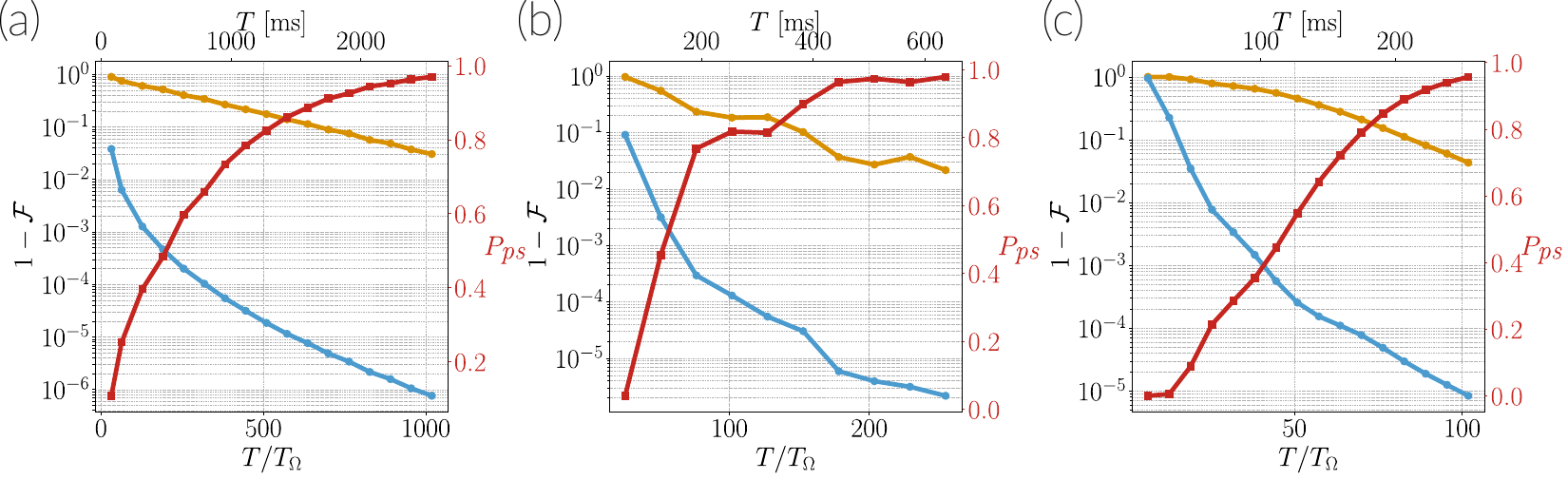}
    \caption{Infidelity before (orange) and after (blue) post-selection on the atom number $N_t=8$, and post-selection probability (red) for the generation of (a) the Laughlin state for a $C^0$ ramp with $\Delta_f = -\Delta_0 = 0.1\Omega$ and $F_{\text{max}}=0.1\Omega$, (b) the Laughlin state with an added quartic potential with $V_0=0.001\hbar \Omega$, and (c) the quasi-hole state with a plug potential with $V_0 = 50.0\hbar \Omega$, $\sigma = 0.2l_B$. The parameters are the same used in the main text.}
    \label{fig:post_sel}
\end{figure}

We note that in an actual experiment, one would need to post-select samples whose number is the target $N_t$. Given the probability $P_{\text{ps}}$ that the state occupies the number sector $N_t$, \cref{fig:post_sel}(h) shows the fidelity after post-selection, defined as $\mathcal{F}_{\text{ps}}=\mathcal{F}/P_{\text{ps}}$. This leads to extremely small infidelities with even faster ramp times, but with a sizeable post-selection probability, as the main effect of diabaticity is the population of the wrong number sector.

\end{document}